\documentclass[preprint,prd,showpacs,amssymb,floatfix,10pt]{revtex4}
\usepackage{graphicx}
\usepackage[latin1]{inputenc}
\usepackage{amsmath}
\usepackage{amsfonts}
\usepackage{amssymb}
\usepackage{amsthm}
\usepackage{bm}
\usepackage{subfigure}
\usepackage{mathrsfs} 

\begin{document}

\title{\bf DeWitt-Schwinger Renormalization and Vacuum Polarization in 
$d$ Dimensions}
\author{R. T. Thompson}
\email{rthompson@fisica.ist.utl.pt}
\affiliation{Centro Multidisciplinar de Astrof\'{\i}sica - CENTRA \\
Departamento de F\'{\i}sica, Instituto Superior T\'ecnico - IST,\\
Universidade T\'ecnica de Lisboa - UTL,\\
Av. Rovisco Pais 1, 1049-001 Lisboa, Portugal \\
\&\\
Department of Mathematics and Statistics,
University of Otago, \\
P.O. Box 56, 
Dunedin, 9054,  New Zealand}
\author{Jos\'e P. S. Lemos}
\email{joselemos@ist.utl.pt}
\affiliation{Centro Multidisciplinar de {}Astrof\'{\i}sica - CENTRA \\
Departamento de F\'{\i}sica, Instituto Superior T\'ecnico - IST,\\
Universidade T\'ecnica de Lisboa - UTL,\\
Av. Rovisco Pais 1, 1049-001 Lisboa, Portugal}

\begin{abstract}
Calculation of the vacuum polarization, $\langle\phi^2(x)\rangle$, and
expectation value of the stress tensor, $\langle T_{\mu\nu}(x)\rangle$,
has seen a recent resurgence, notably for black hole spacetimes.  To
date, most calculations of this type have been done only in four
dimensions. Extending these calculations to $d$ dimensions includes
$d$-dimensional renormalization.  Typically, the renormalizing terms
are found from Christensen's covariant point splitting method for the
DeWitt-Schwinger expansion.  However, some manipulation is required to
put the correct terms into a form that is compatible with problems of
the vacuum polarization type.  Here, after a review of the current
state of affairs for $\langle\phi^2(x)\rangle$ and $\langle
T_{\mu\nu}(x)\rangle$ calculations and a thorough introduction to the
method of calculating $\langle\phi^2(x)\rangle$, a compact expression
for the DeWitt-Schwinger renormalization terms suitable for use in
even-dimensional spacetimes is derived.  This formula should be useful
for calculations of $\langle\phi^2(x)\rangle$ and $\langle
T_{\mu\nu}(x)\rangle$ in even dimensions, and the renormalization
terms are shown explicitly for four and six dimensions.  Furthermore,
use of the finite terms of the DeWitt-Schwinger expansion as an
approximation to $\langle\phi^2(x)\rangle$ for certain spacetimes is
discussed, with application to four and five dimensions.
\end{abstract}


\maketitle
\section{Introduction}

\subsection{Vacua and particle creation}
Hawking radiation shows that a black hole and its horizon are
spacetime regions where gravitational effects become important in
quantum field theory.  The initial approach followed by Hawking
\cite{Hawking:1974sw} was to study quantized ingoing and outgoing
field modes in a fixed black hole background.  This process yields a
flux of particles, produced and emitted out of the surrounding quantum
vacuum with a thermal spectrum.  In a flat spacetime the vacuum is
uniquely defined and relatively easy to find, but in a curved
spacetime it is not so simple.  There are cases where it may be
impossible to define a vacuum state, as in strongly time dependent
geometries. On the contrary, there are situations where many vacua
exist for a given geometry.  The Schwarzschild solution, for example,
has three possible vacua: Boulware \cite{Boulware:1974dm}, Unruh
\cite{Unruh:1976db}, and Hartle-Hawking \cite{Hartle:1976tp}, all
three of which are interesting and important. The Boulware vacuum
describes the vacuum in a region near the surface of a highly compact
star, where it has a small negative energy density. Near the horizon
of a black hole, however, this energy density blows up, so the
Boulware vacuum is inappropriate for a black hole geometry.  On the
other hand, the Unruh and Hartle-Hawking vacua are both consistent for
the Schwarzschild black hole.  The Hartle-Hawking vacuum has a small,
finite, negative energy at the event horizon, which in turn is
responsible for the production of particles and subsequent Hawking
radiation at a given temperature.  The Unruh vacuum, also having a
small, finite, negative energy at the event horizon, is produced by
the complete gravitational collapse of an object, and in this sense is
more physical. However, the one that usually simplifies the
calculations is the Hartle-Hawking vacuum, and, if desired, it is
possible to pass from this to the Unruh vacuum via appropriate
transformations.

Once a consistent vacuum for the geometry in question, such as a black
hole spacetime, has been defined, the associated quantities of
interest may be found.  For a given field $\phi(x)$ and vacuum
$\vert0\rangle$, where $x$ represents a spacetime point, these
quantities may be the vacuum expectation value of the field operator
$\phi^2(x)$ i.e.\ $\langle0\vert\phi^2(x)\vert0\rangle$, or
$\langle\phi^2(x)\rangle$ for short, and its associated vacuum
expectation value of the stress-energy tensor $T_{\mu\nu}(x)$ i.e.
$\langle0\vert T_{\mu\nu}(x)\vert0\rangle$, or $\langle
T_{\mu\nu}(x)\rangle$ for short.  The quantity
$\langle\phi^2(x)\rangle$ is a useful tool in the study of quantum
effects in curved spacetimes. When properly renormalized it gives
information about vacuum polarization effects and spontaneous symmetry
breaking phenomena, although since it is a scalar it does not
distinguish between future and past surfaces, such as horizons and
infinities.  The quantity $\langle T_{\mu\nu}(x)\rangle$ provides
information about the energy density and particle production.
Moreover, since in general relativity Einstein's equations relate the
spacetime curvature to the distribution of matter as encoded in the
stress-energy tensor, for quantum fields the expectation value
$\langle T_{\mu\nu}(x)\rangle$ is used to determine how the underlying
geometry responds to suitable averages of the quantum fields.  These
back-reaction effects by the quantum fields on the background
spacetime are described by the semiclassical Einstein equations
$G_{\mu\nu}=8\pi\langle T_{\mu\nu}(x)\rangle$ (we use $G=1$, $c=1$,
$\hbar=1$) (see \cite{Birrell:1982ix,Fulling:1989nb,Wald:1995yp} for
reviews and careful explanations). The problem of quantum back
reaction is certainly significant in the case of black holes and in
other spacetimes with horizons, such as de Sitter spacetime. For
instance, in the black hole case it leads to the complete evaporation
of the black hole.

\subsection{Renormalizing the vacuum}
Since $\langle\phi^2(x)\rangle$ and $\langle T_{\mu\nu}(x)\rangle$ are
constructed with product expressions, bilinear in the field operators
and evaluated at the same spacetime point, the vacuum expectation
value of these quantities diverges.  For a theory to have any physical
meaning it must give finite results, so some process must be employed
to render these quantities finite.  Such a process amounts to
subtracting off some ``unphysical'' infinite terms. In flat spacetime,
standard normal ordering techniques and other procedures in quantum
field theory work well in regularizing and renormalizing the fields,
where regularization identifies the infinities, and renormalization
eliminates them.  In general relativity, however, the energy density
itself is a source of curvature.  Therefore, when working with a
quantum field theory whose energy density is formally divergent we
must be very careful about what may be dismissed as unphysical.  The
standard techniques used in flat spacetimes do not work in curved
spacetimes, so one of the great difficulties in understanding quantum
processes in a black hole -- or any other curved -- spacetime is the
implementation of consistent regularization and renormalization
schemes.

Fortunately, there are several generally accepted consistent,
covariant regularization and renormalization schemes for curved
spacetimes (see e.g.\ \cite{Birrell:1982ix}).  Of these, the most
widely and consistently used method for $\langle\phi^2(x)\rangle$ and
$\langle T_{\mu\nu}(x)\rangle$ calculations is that of isolating the
divergent terms of the DeWitt-Schwinger expansion.  This covariant
geodesic point separation method, developed by Schwinger
\cite{Schwinger:1951nm}, DeWitt \cite{DeWitt:1965jb,DeWitt:1975ys},
and Christensen
\cite{ChristensenThesis,Christensen:1976vb,Christensen:1978yd} (see
Barvinsky and Vilkovisky in Ref.\ \cite{Barvinsky:1985an} for further
information), is now usually called the point splitting method.  The
idea of the point splitting method is that the operator in each
product is moved along a geodesic to a nearby spacetime point. The
point separated object is expressed in terms of Green's functions
$G(x,x')$.  In the coincidence limit, where the nearby spacetime point
$x'$ approaches the original point $x$, there will be terms diverging
logarithmically (in even dimensions) and as inverse powers of the
point separation.  This point splitting method leads naturally to the
DeWitt-Schwinger expansion, which gives an approximation for the
Green's function $G(x,x')$ when the points $x$ and $x'$ are separated
by a small geodesic distance, $s$, along the shortest geodesic
connecting them.  The result is actually expanded in powers of the
field mass $m$, with expansion coefficients $a_k$ expressed in terms
of geometrical quantities constructed from the Riemann tensor.  Other
renormalization methods exist, for example, dimensional continuation
\cite{Brown:1980qq}.  Many of these methods have been shown to be
equivalent to the DeWitt-Schwinger approach \cite{DeWitt:1975ys}.

The divergent terms of the DeWitt-Schwinger expansion are then the
renormalizing counter terms to be subtracted from the unrenormalized
exact expression for $G(x,x')$, prior to taking the limit
$x\rightarrow x'$.  Now the expressions for $\langle\phi^2(x)\rangle$
and $\langle T_{\mu\nu}(x)\rangle$, written conveniently in terms of
Green's functions, are in fact the properly renormalized
expressions $\langle\phi^2(x)\rangle_{\rm ren}$ and $\langle
T_{\mu\nu}(x)\rangle_{\rm ren}$.  These renormalized values are those
which provide information on spontaneous symmetry breaking and
particle production, as well as being essential for calculating the
backreaction by quantum fields on the spacetime. This means that
semiclassical general relativity has physical meaning when described
by the equations $G_{\mu\nu}=8\pi\langle T_{\mu\nu}(x)\rangle_{\rm
ren}$.  With this renormalization process in hand, a complete set of
mode functions and their associated creation and annihilation
operators must be found by solving the field equations for $\phi(x)$.
The vacuum expectation values $\langle\phi^2(x)\rangle_{\rm ren}$ and
$\langle T_{\mu\nu}(x)\rangle_{\rm ren}$ are then put in terms of the
field operators.  In general the result is a sum over products of mode
functions and their derivatives. The sum can, at least in principle,
be performed and a finite result is achieved.

\subsection{DeWitt-Schwinger estimates for $\langle\phi^2(x)\rangle$}
There is an additional pay-off when using the DeWitt-Schwinger
expansion with Christensen's point separation method.  Since the
expansion is in inverse powers of the field mass $m$, it is valid for
many spacetimes provided $m$ is large enough. In this case the finite
terms of the expansion can provide approximations for both
$\langle\phi^2(x)\rangle_{\rm ren}$ and $\langle
T_{\mu\nu}(x)\rangle_{\rm ren}$.  For instance, given a scalar field
$\phi(x)$ the Feynman Green's function corresponds to
$\langle\phi^2(x)\rangle$ (see e.g.\ \cite{Birrell:1982ix}), so that
the finite terms of the expansion directly give a physical
$\langle\phi^2(x)\rangle_{\rm ren}$.  The major obstacle in the
DeWitt-Schwinger expansion is to compute the coefficients $a_k$. For a
scalar field the first three coefficients, $a_0$, $a_1$, and $a_2$,
have been computed by DeWitt \cite{DeWitt:1965jb,DeWitt:1975ys}; the
coefficient $a_3$ has been computed in the coincidence limit by Gilkey
\cite{Gilkey:1975iq}; and, the coefficient $a_4$ has been computed in
the coincidence limit by Avramidi \cite{Avramidi:1990je}, and by
Amsterdamski, Berkin, and, O'Connor \cite{Amsterdamski:1989bt}.
Additionally, Barvinsky et.\ al.\ \cite{Barvinsky:1994ic} have
calculated these coefficients using different methods.  Thus,
Christensen's method plays definitely two roles -- it is the basis for
point splitting renormalization, and it yields an estimate for the
quantities $\langle\phi^2(x)\rangle_{\rm ren}$ and $\langle
T_{\mu\nu}(x)\rangle_{\rm ren}$.

It should be stressed that the DeWitt-Schwinger expansion together
with Christensen's point separation method is an approximation that
does not hold in all regions of all spacetimes.  For example, the
results of Kay and Wald \cite{Kay:1991} show that for a
Reissner-Nordstr\"om black hole in asymptotically de Sitter spacetime,
$\langle\phi^2(x)\rangle$ cannot be regular on both the event and
cosmological horizons when these horizons have unequal temperatures.

\subsection{Calculations of $\langle\phi^2(x)\rangle$ and applications}
Christensen's work is quite general, and in principle can be applied
to any spacetime. For cosmological as well as some black hole
applications, see Ref.\ \cite{Birrell:1982ix} for works up to around
1980. Many other examples can be given. For massless scalar fields
analytical results were reported by several authors. Candelas studied a
massless scalar field minimally coupled in the Schwarzschild geometry,
where $\langle\phi^2(x)\rangle$ and $\langle T_{\mu\nu}(x)\rangle$
were worked out on the event horizon \cite{Candelas:1980zt}.  Candelas
and Howard \cite{Candelas:1984pg} and Fawcett and Whiting
\cite{Fawcett:1981fw} extended the calculation of
$\langle\phi^2(x)\rangle$ to the exterior region. Candelas and Jensen
\cite{Candelas:1985ip} calculated $\langle\phi^2(x)\rangle$ in the
interior region, and finally Howard and Candelas
\cite{Howard:1984qp,Howard:1985yg} and Fawcett \cite{Fawcett:1983dk}
calculated $\langle T_{\mu\nu}(x)\rangle$ for the whole of
Schwarzschild, definitively extending the pioneering work of Candelas
\cite{Candelas:1980zt}.  In this context of Schwarzschild black holes
it was shown by Hawking \cite{Hawking:1980ng} and Fawcett and Whiting
\cite{Fawcett:1981fw} that the mean square field
$\langle\phi^2(x)\rangle$ can give considerable insight into the
physical content of the different possible vacua and in the study of
theories with spontaneous symmetry breaking.  Massless scalar fields
in Reissner-Nordstr\"om and Kerr-Newman spacetimes were studied by
Frolov \cite{Frolov:1982pi}, where $\langle\phi^2(x)\rangle$ was found
on the event horizon of a Reissner-Nordstr\"om black hole and on the
pole of the event horizon of a Kerr-Newman black hole.  Since
analytical and numerical calculations are difficult, approximation
schemes have been devised for calculating $\langle\phi^2(x)\rangle$
and $\langle T_{\mu\nu}(x)\rangle$ for Schwarzschild,
Reissner-Nordstr\"om, and Kerr-Newman black holes
\cite{Page:1982fm,Brown:1986jy,Zannias:1984tb,Frolov:1987gw,
Frolov:1989jh}.  For massless electromagnetic fields several works
have calculated $\langle T_{\mu\nu}(x)\rangle$ for Schwarzschild,
Reissner-Nordstr\"om, and Kerr-Newman black holes
\cite{Elster:1984hu,Frolov:1984ra,Jensen:1988rh,Matyjasek:1996ih}.
All these works are for massless fields, where calculations may
simplify due to the conformal invariance of the system.

A new approach came with the work of Anderson
\cite{Anderson:1989vg,Anderson:1990jh} where the method was applied
consistently to massive scalar fields.  In Ref.\
\cite{Anderson:1989vg} $\langle\phi^2(x)\rangle$ was calculated for a
generically coupled massive scalar field in the Schwarzschild
geometry, and in \cite{Anderson:1990jh} a powerful formalism was laid
down for finding $\langle\phi^2(x)\rangle$ in a general static
spherical geometry, which includes Schwarzschild and
Reissner-Nordstr\"om solutions.  This was possible through the use of
the Plana sum formula which converts sums into integrals. The method
was extended by Anderson, Hiscock, and Samuel
\cite{Anderson:1993if,Anderson:1994hg} to find $\langle
T_{\mu\nu}(x)\rangle_{\rm ren}$ for a massive scalar field in a
general, static, spherical geometry.  The approach of Refs.\
\cite{Anderson:1990jh,Anderson:1993if,Anderson:1994hg} uses a 
Wentzel-Kramers-Brillouin (WKB)
approximation for the mode functions to compute
$\langle\phi^2(x)\rangle$ and $\langle T_{\mu\nu}(x)\rangle$ to orders
$m^{-4}$ and $m^{-2}$ respectively.  It is further found that, when
applied to the Reissner-Nordstr\"om spacetime, the DeWitt-Schwinger
expansion provides values quite close to the numerical results when
the field mass $mM \gtrsim 1$, where $M$ is the black hole mass (we
put $\hbar=1$).  Anderson's approach has been developed and applied to
other cases.  Cylindrical black hole spacetimes have been examined by
DeBenedictis \cite{DeBenedictis:1998be}, who worked out
$\langle\phi^2(x)\rangle$ for scalar fields, and by Piedra and Oca,
who have studied spinor fields \cite{Piedra:2007yi}. Sushkov
\cite{Sushkov:2000me} studied it for wormholes, Berej and Matyjasek
\cite{Berej:2002xd} for the spacetime of a nonlinear black hole, Satz,
Mazzitelli, and Alvarez \cite{Satz:2004hf} for the vacuum outside
stars, and Winstanley and Young \cite{Winstanley:2007tf} for lukewarm
black holes.  Finally, Flachi and Tanaka \cite{Flachi:2008sr} have
used Anderson's method to compute $\langle\phi^2(x)\rangle$ in
asymptotically anti-de Sitter black hole geometries.  New approaches
have been devised by Anderson, Mottola, and Vaulin
\cite{Anderson:2007eu}, whereas Popov and Zaslavskii have discussed
the WKB approximation in the massless limit \cite{Popov:2007ib}.

\subsection{Renormalization in $d$ dimensions and this paper}

Christensen has remarked \cite{Christensen:1978yd} that while his
methods are valid in arbitrary dimensions, where the procedure is the
same as in four dimensions, calculating quantities such as
$\langle\phi^2(x)\rangle$ and $\langle T_{\mu\nu}(x)\rangle$ in higher
dimensions ``would be extremely long and would probably have to be
done on a computer,'' mainly due to the complexity of the
renormalization problem. This comment is still true and the very few
works since 1978 that have tried to come to terms with the
renormalization techniques in curved $d$-dimensional spacetimes do
prove the difficulty of the extension of the procedure.  Nevertheless
the interest in these techniques to spacetimes with more than four
dimensions has been renewed as the result of progress in areas such as
string theory, AdS/CFT (anti-de Sitter/conformal field theory)
conjecture, Kaluza-Klein theories, extra large-dimensional scenarios,
and the related brane world scenarios.

Earlier, Frolov, Mazzitelli, and Paz \cite{Frolov:1989rv} studied
polarization effects in black hole spacetimes in higher dimensions.
In the context of black holes in a braneworld, Casadio
\cite{Casadio:2003jc} discussed back reaction issues.  In a very
thorough work Decanini and Folacci \cite{Decanini:2005gt} expressed
the DeWitt-Schwinger representation of the Feynman propagator as a
Hadamard expansion for even and odd dimensions which clearly exhibit
the divergent and the regular parts of the DeWitt-Schwinger
representation.  In \cite{Decanini:2005eg,Decanini:2007gj} these
authors presented the first explicit calculations of the stress-energy
tensor in an arbitrary spacetime of $d$ dimensions and provided an
expression for $d=6$ in the large mass limit.  Following the ideas
developed in Christensen and Fulling \cite{Christensen:1977jc},
Morgan, Thom, Winstanley, and Young \cite{Morgan:2007hp} have worked
out some properties of $\langle T_{\mu\nu}(x)\rangle$ for
$d$-dimensional spherical black holes.  Herdeiro, Ribeiro, and Sampaio
\cite{Herdeiro:2007eb} studied the scalar Casimir effect on a
$d$-dimensional Einstein static universe where renormalization
techniques are also used and where, incidentally, the Plana sum
formula (also called the Abel-Plana formula) has been applied -- in
fact the formula was used for the first time in the context of
renormalization techniques for the Casimir effect by Mamaev,
Mostepanenko, and Starobinsky \cite{Mamaev:1976je}; see the review
\cite{Bordag:2001qi}.

In this paper we use the techniques developed by DeWitt
\cite{DeWitt:1975ys}, Christensen
\cite{ChristensenThesis,Christensen:1976vb,Christensen:1978yd}, and
Anderson \cite{Anderson:1990jh,Anderson:1993if,Anderson:1994hg} (see
also \cite{Candelas:1980zt,Candelas:1984pg}) and apply them to the
problem of renormalization of the divergent quantity
$\langle\phi^2(x)\rangle$ for a massive scalar field $\phi(x)$ in a
$d$-dimensional static spacetime, carrying out the renormalization by
the point splitting technique. The derivations presented by DeWitt
\cite{DeWitt:1975ys} and Christensen
\cite{ChristensenThesis,Christensen:1976vb,Christensen:1978yd} are
quite mathematical in nature, and the end result is not in a form that
is amenable for renormalizing $\langle\phi^2(x)\rangle$. The purpose
of this paper is to present a compact formula for the renormalization
terms that may be applied to $\langle\phi^2(x)\rangle$ calculations,
which we achieve for even dimensions.  As applications of our results
in even dimensions, we single out $d=4$ and $d=6$.  In $d=4$ we
compare our results with previous results, and surely, it is the most
important dimension. We then have chosen $d=6$ both because it is the
simplest case after $d=4$ and can be
consistently realized if one advocates extra large dimension or
braneworld scenarios.  Odd dimensions may require other methods to
find a compact formula for the renormalization terms.  In the
calculation we also find $\langle\phi^2(x)\rangle_{\rm ren}$ in first
approximation for large enough field masses, in both even and odd
dimensions. We give as examples the cases $d=4$ and $d=5$. Again,
$d=4$ is singled out because it is the most important dimension and it
can be compared immediately with the previous results of other
authors, and $d=5$ is the first odd higher dimension, and could as
well be important in scenarios with large extra dimensions In brief,
there are two purposes: one is to kill the divergences in
$\langle\phi^2(x)\rangle_{\rm ren}$; the other is to extract the
finite part of $\langle\phi^2(x)\rangle_{\rm ren}$.

The paper is organized as follows.  In Sec.\ \ref{sec:Polarization}
the calculation of $\langle\phi^2(x)\rangle$ is thoroughly reviewed,
including a discussion of the connection between Green's functions and
operator theory, and an outline of the standard method for computing
$\langle\phi^2(x)\rangle$ in a static spacetime.  This motivates the need
for finding a compatible expression for the renormalization terms and shows
what form they must take.  In Sec.\ \ref{sec:SDRenormalization} the
DeWitt-Schwinger expansion for $d$ dimensions is presented, and
isolation of the divergent terms is reviewed.  In Sec.\
\ref{sec:EvenDimensions} even dimensions are studied. 
Specifically, in Sec.\ \ref{sec:Generaltreatment2}, 
an integral representation for the modified
Bessel function $K_{\nu}(z)$ in the limit of vanishing $z$ is
derived for even-dimensional spacetimes.  For scalar fields of zero
temperature, this integral representation may be used in the
expression for the divergent terms.  For a scalar field at temperature
$T$, further manipulation is required to make the expression for the
divergent terms useful for $\langle\phi^2(x)\rangle$ calculations.  The
Plana sum formula is employed to convert the integral into a sum plus
residual terms, leading to a suitable formula for the renormalization
terms in the nonzero temperature case.  As an example, the
renormalization terms are found for four- and six-dimensional
spacetimes in Sec.\ \ref{sec:Examples}.  Section
\ref{sec:Estimates} discusses estimating $\langle\phi^2(x)\rangle$ from
the finite terms of the DeWitt-Schwinger expansion, and some concrete
examples are given for scalar fields in four- and five-dimensional
black hole spacetimes. The results are summarized in Sec.\
\ref{sec:Conclusions}. In the Appendices \ref{App:HowardIdentities}  and 
\ref{App:IntegralRep} we develop some formulas 
needed in the main part of the work.

\section{Vacuum Polarization in $d$-Dimensional Static Spacetimes} 
\label{sec:Polarization}
\subsection{Green's Function Connection to $\langle\phi^2(x)\rangle$}

For a scalar field $\phi(x)$ in a curved spacetime background we
start with the action 
\begin{equation} \label{Eq:actionphi}
S=\int d^dx \sqrt{|g|}{\cal L}\,,
\end{equation}
where $g$ is the determinant of the 
$d$-dimensional spacetime metric, 
 and ${\cal L}$ is the Lagrangian for the scalar field 
$\phi$, given by
\begin{equation} \label{Eq:lagrangianphi}
{\cal L}=\frac12
\left\{g^{\mu\nu}(x)\phi(x)_{,\mu}\phi(x)_{,\nu}
-\left[m^2+\xi\,R(x)\right]\phi^2(x)\right\}.
\end{equation}
Here $m$ is the mass of the field quanta, 
and it is assumed there is a coupling between the scalar and gravitational 
fields of the form $\xi\,R(x)\phi^2(x)$,
where $\xi$ is the coupling constant and $R(x)$ is 
the Ricci scalar of the background spacetime. 
Minimal coupling corresponds to $\xi=0$, while for 
$\xi=\frac14\frac{d-2}{d-1}$ ($\xi=\frac16$ in $d=4$) 
the field is conformally coupled when $m=0$, i.e.\ 
the action is invariant under 
conformal transformations of the type 
$g_{\mu\nu}(x)\rightarrow{\bar g_{\mu\nu}(x)}
=\Omega(x)^2g_{\mu\nu}(x)$ and 
$\phi(x)\rightarrow {\bar \phi(x)}=\Omega(x)^{(2-d)/2}\phi(x)$.
Varying the action of Eq.\ (\ref{Eq:lagrangianphi}) in relation to
$\phi$ gives the equation of motion 
for the field,
\begin{equation} \label{Eq:WaveEqforphi}
\left(\square+ m^2 + \xi R(x)\right)\phi(x)=0\,.
\end{equation}
This is a generalized covariant Klein-Gordon equation, 
where $\square=g^{\mu\nu}\nabla_\mu\nabla_\nu\, 
=|g|^{-1/2}\left[(|g|)^{1/2}g^{\mu\nu}\phi_{\,\nu}\right]_{,\mu}$ 
is the Laplace-Beltrami operator
in $d$-dimensional curved spacetime.

Quantization reveals that the field is composed
of particles obeying certain commutation relations, and one wants
to know how these particles move or propagate in the given curved
background spacetime.  A propagator is usually defined by modifying
the Klein-Gordon equation, Eq.\ (\ref{Eq:WaveEqforphi}), so as to include a
source term $J(x)$ such that $\left(\square+ m^2 + \xi
R(x)\right)\phi(x)=J(x)$. This equation may be solved using the
standard theory of Green's functions. One of the Green's functions, or
propagators, that can be defined is $G_{\rm F}(x,y)$, which satisfies
\begin{equation} \label{Eq:WaveEqforgreensfunction} \left(\square+ m^2
+ \xi R(x)\right) G_{\rm F}(x,x')= -|g(x)|^{-1/2}\delta^d(x-x')\,,
\end{equation} 
where $\delta(x)$ is the Dirac delta function. The
solution for $\phi(x)$ is then $\phi(x)=\phi_0(x)-\int d^d x\, 
|g(x)|^{1/2} G_{\rm
F}(x,x')\,J(x')$, where $\phi_0(x)$ is a function that satisfies the
Klein-Gordon equation without a source term and 
$\phi(x)$ corresponds to a quantum field 
operator acting on some state.

Interestingly,
vacuum expectation values of products of field operators
can be identified with the Green's function of the wave equation, 
as we now show. 
The propagation of a free test particle in a 
vacuum $\vert0\rangle$ can be described by the correlation 
function $G^+(x,x')=\langle0\vert\phi(x)\phi(x')\vert0\rangle
\equiv \langle\phi(x)\phi(x')\rangle$, 
where $\phi(x')$ creates a particle at $t'$, which in 
turn is annihilated by $\phi(x)$ at $t$. This makes sense if 
$t>t'$. Analogously, 
the correlation 
function $G^-(x,x')=\langle\phi(x')\phi(x)\rangle$, 
describes the propagation of a particle 
created by $\phi(x)$ at time $t$, which in 
turn is annihilated by $\phi(x')$ at $t'$. This makes sense if 
$t'>t$. To obtain a correlation function, or propagator, 
that has physical meaning in relativistic quantum field theory, 
either $G^+(x,x')$ or $G^-(x,x')$ is used, depending 
on the sign of the relative time. 
So to obtain a physically meaningful propagator that 
combines both $G^+(x,x')$ and $G^-(x,x')$
we can use $\langle T\left(
\phi(x)\phi(x')\right)\rangle$, 
where Dyson's time ordering product operator $T$ is defined as 
$T\left(\phi(x)\phi(x')\right)=
\theta(t-t')\phi(x)\phi(x')+\theta(t'-t)\phi(x')\phi(x)$, 
with $\theta(t)=1$ for $t>0$ and $\theta(t)=0$ for $t<0$.  To call
$T\left(\phi(x)\phi(x')\right)$ the ``time ordered product'' is apt
since the operators occurring under the symbol $T$ are arranged from
right to left with increasing times.  Such a propagator is called the
Feynman propagator, and one can show that this time ordered product of
fields is indeed the Feynman Green's function defined by Eq.\
(\ref{Eq:WaveEqforgreensfunction}), i.e.
\begin{equation} \label{Eq:Fgreenfunction}
 iG_{\rm{F}}(x,x') = \langle T \left(\phi(x)\phi(x')\right) 
 \rangle.
\end{equation}
Using the Klein-Gordon equation, Eq.\
(\ref{Eq:WaveEqforphi}), and the properties 
of the step function $\theta(t'-t)$, one finds
\begin{equation}
\left(\square_x+ m^2 + \xi R(x)\right)\langle T\left(
\phi(x)\phi(x')\right) \rangle=-i|g(x)|^{-1/2}\delta^d
(x-x').
\end{equation}
Care should be taken since the step functions are time-dependent, and
instead of zero the result is a distribution $\delta^d (x-x')$
concentrated at equal times.  Thus it follows that the vacuum
expectation value $\langle T\left( \phi(x)\phi(x')\right) \rangle$ is
essentially one of the Green's functions of the covariant generalized
Klein-Gordon operator, and we are justified in calling it the Feynman
Green's function $G_{\rm F}(x,x')$.  In other words, the analysis
shows that the Feynman propagator is a Green's function of the
Klein-Gordon equation.

Usually in quantum field theory 
the equation connecting Green's functions and expectation values, 
such as Eq.\ (\ref{Eq:Fgreenfunction}), 
gives a bridge between the theory of propagators, in which
scattering amplitudes are written in term of Green's functions, 
and the theory of operators, where everything is written in terms of 
the quantum field $\phi(x)$. One finds
the operators and expectation values, thus obtaining the Green's 
functions important for interaction theory. 
We see that in our study, the 
connection is inverted -- we want $\phi^2(x,x')$ at the point
$x$ by expressing operator theory in terms of the Green's function and so
we calculate the Green's function. Thus, Eq.\ 
(\ref{Eq:Fgreenfunction}) operates as a kind of duality.

Since we are interested in the coincidence limit, 
Feynman's Green function is the best to use because 
it is more physical and also because the 
boundary conditions allow a Wick rotation of the equation to 
Euclidean space, where
\begin{equation} \label{Eq:wickfeynman}
G_{\rm F}(t,x;t',x')=-iG_{\rm E}(i\tau,x;i\tau',x').
\end{equation}
The Euclidean Green's function, $G_{\rm E}$, now obeys
\begin{equation} \label{Eq:Fgreenfunctionequationeuclidean}
\left(\square_{\mathrm{E}}- m^2 - \xi R(x)\right)G_{\rm E}(x,x')=
-|g(x)|^{-1/2}\delta^d(x-x')\,.
\end{equation}
where $\square_{\mathrm{E}}$ is now the
Laplace-Beltrami operator
in $d$-dimensional curved Euclidean space.
There are advantages to working in Euclidean space. 
For instance,  elliptic operators are more easily handled than
hyperbolic operators, and after obtaining the Euclidean results 
one can Wick rotate back to Lorentzian spacetime using Eq.\
(\ref{Eq:Fgreenfunctionequationeuclidean}) since the boundary 
conditions for the Feynman propagator are automatically 
imposed by this procedure. 

The Feynman Green's function, or alternatively the Euclidean Green's 
function,
is defined in terms of expectation values of products of field operators 
in the pure vacuum state. This is fine for describing the 
system at zero temperature.  To go further and 
describe a system at nonzero temperature one has to take into 
account that the system is no longer in a pure state, 
it is statistically distributed over all possible states. The full 
weight of statistical physics must be used, and the Green's functions 
are given by the average, suitably weighted, 
over all pure states of the expectation value of the products 
of field operators in those pure states 
(see e.g.\ Ref.\ \cite{fetter}).

\subsection{Calculating the Green's Function}
The standard approach now used for calculating $\langle \phi^2(x)\rangle$
was laid down by Anderson \cite{Anderson:1990jh}, based on
earlier works by Candelas and Howard
\cite{Candelas:1980zt,Candelas:1984pg}.  We start with the Euclidean
metric for a static 
spacetime in $d$ dimensions with line element 
\begin{equation} \label{Eq:Metric}
 ds^2 = f(r)d\tau^2 + h(r)dr^2 + r^2d\Omega^2.
\end{equation}
Here $\tau$ is the Euclidean time, $\tau = -it$, 
$r$ is a kind of radial coordinate, and $\Omega$ represents a 
$(d-2)$-dimensional
angular space.  The only restriction for this method is that the 
metric must be diagonal.  The expectation value $\langle\phi^2(x)\rangle$ 
is found
from the coincidence limit of the Euclidean Green's function
$G_{\rm E}({x},{x})\equiv \lim_{x'\rightarrow x}G_{\rm E}({x},{x'})$.  For a
scalar field in a spacetime given by Eq.\ (\ref{Eq:Metric}),
$G_{\rm E}({x},{x'})$ satisfies [see Eq.\
(\ref{Eq:Fgreenfunctionequationeuclidean})]
\begin{equation} \label{Eq:WaveEq}
\left(\square_{\rm E} - m^2 - \xi R(x)\right)G_{\rm E}({x},{x'}) =
-\frac{1}{\sqrt{g(x)}}\delta(\tau-\tau')\delta(r-r')
\delta(\Omega,\Omega'),
\end{equation}
where $\square_{\rm E}$ is the Laplace-Beltrami operator of the
Euclidean metric corresponding to Eq.\ (\ref{Eq:Metric}).  Assuming a
separation of variables, the independent homogeneous equations for
$\tau$ and $\Omega$ may be solved. Standard Green's function
techniques then tell us that the $\tau$ and $\Omega$ dependence of
$G_{\mathrm{E}}(x,x')$ is equivalent to a representation of the
corresponding delta function.  We therefore use
\begin{equation} \label{Eq:DeltaOmega}
 \delta(\Omega,\Omega')=\sum_{\ell}\sum_{\{\mu_j\}}
Y_{\ell,\{\mu_j\}}(\Omega)Y^*_{\ell,\{\mu_j\}}(\Omega')\,,
\end{equation}
as the \textit{Ansatz} for the angular dependence of the
Euclidean Green's function.  The function $Y_{\ell,\{\mu_j\}}$ in Eq.\
(\ref{Eq:DeltaOmega}) has been generalized to the set of
hyperspherical harmonics.  In four dimensions these are the usual
spherical harmonics such that
$\sum_{\ell}\sum_{m}Y_{\ell,m}(\Omega)Y^*_{\ell,m}(\Omega') = 
\frac{1}{4\pi}\sum_{\ell}(2\ell+1)P_{\ell}(\Omega\cdot\Omega')$.
As the \textit{Ansatz} for the time dependence of the
Euclidean Green's function, an integral or a sum 
representation is used depending on whether the scalar field is 
at zero or nonzero temperature, respectively. 
If the scalar field is at zero temperature, then
\begin{equation} \label{Eq:DeltaTau}
 \delta(\tau-\tau') = \frac{1}{2\pi}\int_{-\infty}^{\infty}
d\omega e^{i\omega(\tau-\tau')}\,.
\end{equation}
If the scalar field is at nonzero temperature $T$, then the Green's function 
is periodic in $\tau-\tau'$ with period $T^{-1}$ \cite{fetter}, 
and a suitable representation for the delta function is
\begin{equation} \label{Eq:DeltaTaunonzero}
 \delta(\tau-\tau') = T\,\sum_{n=-\infty}^{\infty}
\exp\left[in2\pi T(\tau-\tau')\right]\,.
\end{equation}
Henceforth denote $\varepsilon = \tau-\tau'$.  If the
scalar field is at zero temperature, then
\begin{equation} \label{Eq:GreenFuncZeroTemp}
 G_{\mathrm{E}}({x},{x'}) = \frac{1}{2\pi}\int_{-\infty}^{\infty} 
d\omega e^{i\omega\varepsilon}\sum_{\ell}\sum_{\{\mu_j\}}
Y_{\ell,\{\mu_j\}}(\Omega)Y^*_{\ell,\{\mu_j\}}(\Omega'),
\chi_{\omega\ell}(r,r').
\end{equation}
where $\chi_{\omega\ell}(r,r')$ is the last component of the variable 
separated Green's function -- a radial mode function.
On the other hand, for a scalar field at some nonzero temperature $T$
\begin{equation} 
\label{Eq:GreenFuncNonZeroTemp}
 G_{\rm E}({x},{x'}) = \frac{\kappa}{2\pi}
\sum_{n=-\infty}^{\infty} e^{i\kappa\varepsilon n}\sum_{\ell}
\sum_{\{\mu_j\}}Y_{\ell,\{\mu_j\}}(\Omega)Y^*_{\ell,\{\mu_j\}}(\Omega') 
\chi_{n\ell}(r,r'),
\end{equation}
where $\kappa = 2\pi T$.  In both cases the radial function obeys a
differential equation obtained by putting the above expressions into
Eq.\ (\ref{Eq:WaveEq}).  Using this expression for the Green's
function, and the preceding discussion on the connection between the
Green's function and the expectation value $\langle\phi^2(x)\rangle$
allows us to calculate $\langle\phi^2(x)\rangle$ in the Hartle-Hawking
vacuum.

There are three difficulties when evaluating these expressions in the
coincidence limit. The first is that the equation of motion for the
radial function $\chi_{n\ell}$ (equivalently $\chi_{\omega\ell}$) is
quite complicated, with exact solutions only available for zero frequency.
Asymptotic solutions are obtainable in closed form for massless fields
on the horizon of Schwarzschild and Reissner-Nordstr\"om black holes
\cite{Candelas:1980zt,Frolov:1989rv,Anderson:1990jh}.  Partially
analytical and numerical evaluations of the radial modes occupy the
bulk of current research on this topic and will not be discussed here.
The other two difficulties are that the sums over both $\ell$ and $n$
(equivalently the integral over $\omega$) produce divergences. The
divergence resulting from the sum over $\ell$ is actually only an
apparent divergence and may be easily remedied.  The standard trick is
to realize that, given the delta function dependence, we are free to
add a term proportional to the delta function.  The large $\ell$
contributions are then eliminated with the help of a WKB
approximation.  It is the divergence resulting from the sum over $n$
which is more serious, and it is to this matter that we direct our
attention.

\section{DeWitt-Schwinger Renormalization in $d$ Dimensions}
\subsection{General treatment}\label{sec:SDRenormalization}
To assign a physical meaning to $\langle\phi^2(x)\rangle$, it must be
rendered finite via some renormalization process.  The divergence
resulting from the sum over $n$ in Eqs.\ (\ref{Eq:GreenFuncZeroTemp})
and (\ref{Eq:GreenFuncNonZeroTemp}) is related to the high frequency
behavior of the scalar field.  The high frequency modes of the field
probe the spacetime geometry in a small neighborhood of an event.
Since the metric changes negligibly in this neighborhood an adiabatic,
short distance approximation for the propagator should give the same
divergent behavior as Eqs.\ (\ref{Eq:GreenFuncZeroTemp}) and
(\ref{Eq:GreenFuncNonZeroTemp}). Isolating the ultraviolet divergences
with such an approximation, these divergent terms can then be
subtracted from Eqs.\ (\ref{Eq:GreenFuncZeroTemp}) and
(\ref{Eq:GreenFuncNonZeroTemp}); leaving the renormalized, finite part
of the Green's function. The now standard approach is to renormalize
the expression for $G_{\rm E}({x},{x'})$ via the point splitting
method of Christensen applied to the DeWitt-Schwinger expansion of the
propagator
\cite{Christensen:1976vb,Christensen:1978yd,Schwinger:1951nm,DeWitt:1975ys}.
In $d$ dimensions, the adiabatic DeWitt-Schwinger expansion of the
Euclidean propagator is \cite[Eq.\ 3.10]{Christensen:1978yd}
\begin{equation} \label{Eq:SchwingDeWitt}
G_{\rm E}^{\rm DS}({x},{x'}) = \frac{\pi \triangle^{1/2}}{(4\pi
i)^{d/2}} \sum_{k= 0}^{\infty} a_k({x},{x'})
\left(-\frac{\partial}{\partial m^2}\right)^k 
\left(-\frac{z}{2im^2}
\right)^{1-d/2} H^{(2)}_{d/2-1}(z).
\end{equation}
This equation is slightly different than that found in Ref.\
\cite{Christensen:1978yd}, where the expression is given for the
Feynman Green's function rather than the Euclidean Green's function
and uses a different sign convention; the two are related by Eq.\
(\ref{Eq:wickfeynman}).  Equation (\ref{Eq:SchwingDeWitt}) introduces
several new variables that must be defined.
Let $s({x},{x'})$ be the geodesic distance between
${x}$ and ${x'}$, then define
$2\sigma({x},{x'}) = s^2({x},{x'})$ and $z^2=-2m^2\sigma({x},{x'})$.  
The coefficients
$a_k({x},{x'})$ are generally referred to as
DeWitt coefficients.  The function $H^{(2)}_{\nu}(z)$ is a
Hankel function of the second kind. Lastly,
$\triangle({x},{x'}) =
\sqrt{g({x})}D({x},{x'})\sqrt{g({x'})}$ is
the Van Vleck--Morette determinant, where
$g({x})=\det(g_{\mu\nu}({x}))$ and
$D({x},{x'})=\det(-\sigma_{;\mu\nu'})$.
Expressing
$-2m$ and $dm^2$ in terms of $z$ and $dz$ (for fixed $\sigma$), Eq.\
(\ref{Eq:SchwingDeWitt}) can be written as
\begin{equation} \label{Eq:SchwingDeWitt2}
G_{\rm E}^{\rm DS}({x},{x'}) = \frac{-i\pi
\triangle^{1/2}}{(4\pi)^{d/2}} \sum_{k= 0}^{\infty}
a_k({x},{x'})(-2m^2)^{(d/2-1-k)} z^{-2(d/2-1-k)}
\left(\frac{\partial}{z\partial z}\right)^k z^{d/2-1}
H^{(2)}_{d/2-1}(z).
\end{equation}
By the derivative formula for Bessel functions \cite{Abramowitz},
\begin{equation}
\left(\frac{\partial}{z\partial z}\right)^k z^{\mu} H^{(2)}_{\mu}(z) =
z^{\mu-k}H^{(2)}_{\mu-k}(z),
\end{equation}
and defining $\nu=d/2-1-k$, Eq.\ (\ref{Eq:SchwingDeWitt2}) becomes
\begin{equation}
G_{\rm E}^{\rm DS}({x},{x'}) = \frac{-i\pi \triangle^{1/2}}{(4\pi)^{d/2}} 
\sum_{k= 0} a_k({x},{x'}) (-2m^2)^{\nu} z^{-\nu} H^{(2)}_{\nu}(z).
\end{equation}
The idea here is
that the DeWitt-Schwinger expansion results from
a WKB expansion for the Euclidean 
(or Feynman) propagator for a generic
spacetime when the point separation is small.  For a particular
spacetime this procedure does not give the correct results for the
Green's function with finite point separation, but it should 
reproduce the same divergent terms in the coincidence limit.
Therefore, if the divergent terms of the DeWitt-Schwinger expansion
can be isolated, then these will be the terms to subtract from
$G_{\rm E}({x},{x'})$ in order to make it finite as $x\rightarrow x'$.

The Hankel function is related to the usual Bessel functions by
$H^{(2)}_{\nu}(z) = J_{\nu}-iY_{\nu}(z)$.  Note that $z=i|z|$ is
purely imaginary in Euclidean space.  For a purely imaginary argument
one finds \cite{Abramowitz}
\begin{equation}
H^{(2)}_{\nu}(i|z|)=J_{\nu}(i|z|)-iY_{\nu}(i|z|) = i^{\nu}I_{\nu}(|z|)
- i\left[i^{\nu+1}I_{\nu}(|z|) - \frac{2}{\pi}(-i)^{\nu}K_{\nu}(|z|)
\right] = 2i^{\nu}I_{\nu}(|z|)+\frac{2}{\pi}i(-i)^{\nu}K_{\nu}(|z|),
\end{equation}
leading to
\begin{equation}
G_{\rm E}^{\rm DS}({x},{x'}) = \frac{-2i
\triangle^{1/2}}{(4\pi)^{d/2}} \sum_{k= 0} a_k({x},{x'})(2m^2)^{\nu}
|z|^{-\nu}\Big[(-1)^{\nu}\pi I_{\nu}(|z|)+i K_{\nu}(|z|)\Big].
\end{equation}
Since we are working in Euclideanized space the physical
renormalization terms come from the real part of this expression,
which will leave only the $K_{\nu}(|z|)$ terms.  The asymptotic
behavior of $K_{\nu}(|z|)$ for small argument, $z\to 0$, is
\begin{equation}
 K_{\nu}(|z|) \sim
 \begin{cases}
  \left(\frac{2}{|z|}\right)^{\nu}, & \nu > 0\,,\\
  \, \ln\left(\frac{|z|}{2}\right)+\gamma, & \nu = 0\,, \\
  \left(\frac{2}{|z|}\right)^{-|\nu|}, & \nu < 0\,,
 \end{cases}
\end{equation}
clearly only those terms of the sum for which $\nu\geq 0$ produce
divergent terms in the coincidence limit.  Let $k_d$ be the largest
integer that is less than or equal to $d/2-1$; for even dimensions
$k_d = (d-2)/2$ while for odd dimensions $k_d=(d-3)/2$.  It is clear
that the divergences arise from those terms for which $k\leq k_d$.
Note however, that the terms for $k>d/2-1$ are also divergent in the
limit of vanishing field mass.  For a massless field, where the
DeWitt-Schwinger formalism is obviously not well defined, the mass
must be replaced in those terms which diverge logarithmically as $m\to
0$ with a parameter $\mu$ which must be fixed by a renormalization
condition, or perhaps, as claimed in \cite{DeWitt:1975ys}, must
ultimately be determined experimentally. Taking the real part of
$G_{\rm E}^{\rm DS}$, the divergent
terms are 
\begin{equation} \label{Eq:GdivTerms} G_{\rm div}({x},{x'})
= \frac{2 \triangle^{1/2}}{(4\pi)^{d/2}} \sum_{k= 0}^{k_d}
a_k({x},{x'}) (2m^2)^{\nu} |z|^{-\nu} K_{\nu}(|z|).
\end{equation}
The Van Vleck--Morette determinant $\Delta^{1/2}$ and the DeWitt
coefficients $a_k({x},{x'})$ both depend on the point separation $z$,
or equivalently $\sigma({x},{x'}) =
\tfrac{1}{2}\sigma^{\rho}\sigma_{\rho}$, where
$\sigma_{\rho}=\sigma_{;\rho}$ \cite{ChristensenThesis}.  Essentially,
$\sigma^{\rho}$ is a vector that points from ${x}$ to ${x'}$ and has
length equal to the distance from ${x}$ to ${x'}$.  Consequently,
$\sigma^{\rho}\to 0$ in the coincidence limit.

Any scalar function may be expanded in a covariant
Taylor series of the form \cite{Barvinsky:1985an}
\begin{equation}
f(x') =
\sum_{k=0}^{\infty}\frac{(-1)^k}{k!}\nabla_{\alpha_1}\dots\nabla_{\alpha_k}
f(x) \sigma^{\alpha_1}\dots\sigma^{\alpha_k}.
\end{equation}
Christensen has calculated these expansions for $\Delta^{1/2}$, $a_0(x,x')$,
$a_1({x},{x'})$, and $a_2({x},{x'})$,
which are provided here for completeness \cite{ChristensenThesis,
Christensen:1978yd}.  He finds, with $a_0=1$,
 
 \begin{equation} \label{Eq:akExpansions1}
  \Delta^{1/2} = 1 +
  \tfrac{1}{12}R_{\alpha\beta}\sigma^{\alpha}\sigma^{\beta} -
  \tfrac{1}{24}R_{\alpha\beta;\gamma}\sigma^{\alpha}\sigma^{\beta}
\sigma^{\gamma}
  + (\tfrac{1}{288}R_{\alpha\beta}R_{\gamma\delta} +
  \tfrac{1}{360}R^{\rho\phantom{\alpha}\tau}_{\phantom{\rho}\alpha
\phantom{\tau}\beta}R_{\rho\gamma\tau\delta}+\tfrac{1}{80}
R_{\alpha\beta;\gamma\delta})\sigma^{\alpha}\sigma^{\beta}
\sigma^{\gamma}\sigma^{\delta}+\cdots,
 \end{equation}
\begin{multline}\label{Eq:akExpansions2}
a_1 = (\tfrac{1}{6}-\xi)R -
\tfrac{1}{2}(\tfrac{1}{6}-\xi)R_{;\alpha}
\sigma^{\alpha}\\+\left[-\tfrac{1}{90}
R_{\alpha\rho}R^{\rho}_{\phantom{\rho}\beta}
+\tfrac{1}{180}R^{\rho\tau}R_{\rho\alpha\tau\beta}
+\tfrac{1}{180}R_{\rho\tau\kappa\alpha}
R^{\rho\tau\kappa}_{\phantom{\rho\tau\kappa}\beta}
+\tfrac{1}{120}R_{\alpha\beta;\rho}^{\phantom{\alpha\beta;\rho}\rho}
+(\tfrac{1}{40}-\tfrac{1}{6}\xi)R_{;\alpha\beta}
\right]\sigma^{\alpha}\sigma^{\beta} + \cdots,
 \end{multline}
\begin{equation} \label{Eq:a2}
a_2 = -\tfrac{1}{180}R^{\rho\tau}R_{\rho\tau} +
\tfrac{1}{180}R^{\alpha\beta\rho\tau}R_{\alpha\beta\rho\tau}
+\tfrac{1}{6}(\tfrac{1}{5}-\xi)R_{;\rho}^{\phantom{;\rho}\rho} +
\tfrac{1}{2}(\tfrac{1}{6}-\xi)^2R^2 + \cdots.
 \end{equation}
For calculations of $\langle\phi^2\rangle$ up to $d=7$, only the
DeWitt coefficients up to $a_2$ are required for renormalization.  For
$d=8,9$, $a_3$ is needed and has been found by Gilkey
\cite{Gilkey:1975iq}.  For $d=10,11$ $a_4$ is needed and has been
calculated in the coincidence limit
\cite{Avramidi:1990je,Amsterdamski:1989bt,Barvinsky:1994ic}.  For
higher dimensional spacetimes, subsequent $a_n$ coefficients must be
calculated.  Note however that for calculations of $\langle
T_{\mu\nu}\rangle$ or other quantities involving derivatives of the
field, more of the $a_n$ may be required for a given dimension. Since
$\Delta^{1/2}$ and the $a_k$ contain powers of $\sigma^{\rho}$, the
entire expression for $G_{\rm div}$ should be expanded in powers of
$\sigma^{\rho}$ before taking the coincidence limit ${x}\to{x'}$.  Let
$\Delta^{1/2}$ and $a_k$ be expressed as
 \begin{equation} \label{Eq:DeltaOneHalf}
  \Delta^{1/2} = \Delta^{1/2}_0 + \Delta^{1/2}_1 + \Delta^{1/2}_2 +\cdots,
 \end{equation}
and
 \begin{equation} \label{Eq:akexp}
  a_k = a^0_k + a^1_k + a^2_k + \cdots,
\end{equation}
where the $j^{\mbox{th}}$ numerical index indicates the corresponding
term of Eqs.\ (\ref{Eq:akExpansions1})-(\ref{Eq:a2}) containing $j$
powers of $\sigma^{\rho}$ (note that $\Delta^{1/2}_1=0$).  The bracket
notation is the usual notation found in the literature indicating the
coincidence limit, e.g.\ $[a_k] = a^0_k$ is the term containing zero
powers of $\sigma^{\rho}$.

These expressions can be put together and expanded to the appropriate
order for any dimension. The result, however, would still not be in a
form that can be combined with Eqs.\
(\ref{Eq:GreenFuncZeroTemp})-(\ref{Eq:GreenFuncNonZeroTemp}).  It
would therefore be useful to workers in the field to have a compact
formula for the renormalization terms as applied to calculations of
$\langle\phi^2(x)\rangle$ and $\langle T_{\mu\nu}(x) \rangle$.

The modified Bessel function $K_{\nu}(|z|)$ behaves differently for
even and odd dimensions, so they must be considered separately. In
even dimensions $\nu$ is an integer while for odd dimensions $\nu$ is
a half integer.  In the small $z$ limit one may verify that
$|z|^{-\nu}K_{\nu}(|z|)$ behaves as

\begin{equation} \label{Eq:EvenBesselSmallArg}
 |z|^{-\nu}K_{\nu}(|z|)=\sum_{n=1}^{\nu}
  \frac{(-1)^{\nu-n}\Gamma(n)}{2^{\nu-2n+1}\Gamma(\nu-n+1)|z|^{2n}}+
\frac{(-1)^{\nu}}{2^{\nu+1}\Gamma(\nu
  +1)}\left(\sum_{n=1}^{\nu}
  \frac{1}{n}-2\left(\ln\frac{|z|}{2}+\gamma\right)\right) + O(|z|)
\end{equation}
for integer $\nu$, and as
\begin{equation} \label{Eq:OddBesselSmallArg}
|z|^{-\nu}K_{\nu}(|z|)=\sum_{n=1}^{\nu +\frac{1}{2}}
 \frac{(-1)^{\nu+n+\frac{1}{2}}2^{2n-1}\Gamma(n-1/2)}{2^{\nu+1}\Gamma(\nu
 -n+3/2)|z|^{2n-1}}+\frac{(-1)^{\nu
 +\frac{1}{2}}\pi}{2^{\nu+1}\Gamma(\nu +1)} + O(|z|)
\end{equation}
for half-integral $\nu$.  Incidentally, from these expansions for
$K_{\nu}(|z|)$ one can begin to see the connection with the Hadamard
form of the Green's function \cite{Decanini:2005gt,Decanini:2005eg}.
These expressions imply that
multiplying $\Delta^{1/2}$, $a_k$, and $|z|^{-\nu}K_{\nu}(|z|)$
requires expansions of $\Delta^{1/2}$ and $a_k$ to order $2\nu$ in
$\sigma^{\rho}$ prior to taking the coincidence limit.  Some authors
refer to this as the ``adiabatic order.''  Using the expansions of $a_k$
and $\Delta^{1/2}$, Eqs.\ (\ref{Eq:DeltaOneHalf}) and
(\ref{Eq:akexp}), we may collect $a_k\Delta^{1/2}$ in powers of
$\sigma^{\rho}$,
\begin{equation}
 a_k\Delta^{1/2} = a_k^0\Delta^{1/2}_0 + (a_k^0\Delta^{1/2}_1 + 
a_k^1\Delta^{1/2}_0) + (a_k^0\Delta^{1/2}_2 + a_k^1\Delta^{1/2}_1 + 
a_k^2\Delta^{1/2}_0) + \cdots = [a_k][\Delta^{1/2}] + 
\sum_{p=1}^{\infty}\sum_{j=0}^{p}a^j_k\Delta^{1/2}_{p-j}.
\end{equation}
The summand of Eq.\ (\ref{Eq:GdivTerms}) may be expanded explicitly in
powers of $\sigma^{\rho}$, giving
\begin{multline}
 [a_k][\Delta^{1/2}](2m^2)^{\nu}|z|^{-\nu}K_{\nu}(|z|) +
(2m^2)^{\nu}\sum_{n=1}^{\nu}
\frac{(-1)^{\nu-n}\Gamma(n)}{2^{\nu-2n+1}\Gamma(\nu-n+1)|z|^{2n}}
\sum_{p=1}^{2n}\sum_{j=0}^{p}a^j_k\Delta^{1/2}_{p-j} \\ +
(2m^2)^{\nu}\sum_{n=1}^{\nu}
\frac{(-1)^{\nu-n}\Gamma(n)}{2^{\nu-2n+1}\Gamma(\nu-n+1)|z|^{2n}}
\sum_{p=2n+1}^{\infty}\sum_{j=0}^{p}a^j_k\Delta^{1/2}_{p-j} \\ -
\left[\frac{(-1)^{\nu}}{2^{\nu+1}\Gamma(\nu+1)}
\left[2\left(\ln\frac{|z|}{2}+\gamma\right)-\sum_{n=1}^{\nu}\frac{1}{n} 
\right] + O(|z|^1)\right]\sum_{p=1}^{\infty}\sum_{j=0}^{p}a^j_k
\Delta^{1/2}_{p-j}
\end{multline}
for integral $\nu$ (even dimensions), and
\begin{multline}
 [a_k][\Delta^{1/2}](2m^2)^{\nu}|z|^{-\nu}K_{\nu}(|z|) +
(2m^2)^{\nu}\sum_{n=1}^{\nu+\frac{1}{2}}
\frac{(-1)^{\nu+n+\frac{1}{2}}2^{2n-1}\Gamma(n-1/2)}{2^{\nu+1}
\Gamma(\nu-n+3/2)|z|^{2n-1}}
\sum_{p=1}^{2n}\sum_{j=0}^{p}a^j_k\Delta^{1/2}_{p-j} \\ +
(2m^2)^{\nu}\sum_{n=1}^{\nu+\frac{1}{2}}
\frac{(-1)^{\nu+n+\frac{1}{2}}2^{2n-1}\Gamma(n-1/2)}{2^{\nu+1}
\Gamma(\nu-n+3/2)|z|^{2n-1}}
\sum_{p=2n+1}^{\infty}\sum_{j=0}^{p}a^j_k\Delta^{1/2}_{p-j} +
\frac{(-1)^{\nu+\frac{1}{2}}\pi}{2^{\nu+1}\Gamma(\nu+1)}
\sum_{p=1}^{\infty}\sum_{j=0}^{p}a^j_k
\Delta^{1/2}_{p-j}
\end{multline}
for half-integral $\nu$ (odd dimensions).
In the second term $a_k\Delta^{1/2}$ has been expanded to order
$2\nu$.  Since $a_k^j\Delta^{1/2}_{p-j}$ is proportional to
$(\sigma^{\alpha})^p$, it is clear that the third and subsequent terms
all vanish in the coincidence limit, leaving
\begin{equation} \label{Eq:EvenCovariantRenormTerms}
G_{\rm div}({x},{x'}) = \frac{2}{(4\pi)^{d/2}}\sum_{k=0}^{k_d} \left[
[a_k](2m^2)^{\nu}|z|^{-\nu}K_{\nu}(|z|) +
\sum_{n=1}^{\nu}\sum_{p=1}^{2n}\sum_{j=0}^{p}
\frac{2^{2n-1}(-m^2)^{\nu-n}\Gamma(n)}{\Gamma(\nu-n+1)}\frac{a^j_k
\Delta^{1/2}_{p-j}}{(\sigma^{\rho}\sigma_{\rho})^n}\right],
\end{equation}
and
\begin{equation} \label{Eq:OddCovariantRenormTerms}
G_{\rm div}({x},{x'}) =
\frac{2}{(4\pi)^{d/2}}\sum_{k=0}^{k_d} 
\left[[a_k][\Delta^{1/2}](2m^2)^{\nu}|z|^{-\nu}K_{\nu}(|z|) + 
\sum_{n=1}^{\nu+\frac{1}{2}}\sum_{p=1}^{2n}\sum_{j=0}^{p} 
\frac{2^{2n-2}(-m^2)^{\nu+n+\frac{1}{2}}
\Gamma(n-\frac{1}{2})}{\Gamma(\nu-n+\frac{3}{2})} \frac{a^j_k
\Delta^{1/2}_{p-j}}{(\sigma^{\rho}\sigma_{\rho})^{n-\frac{1}{2}}}\right]
\end{equation}
for even and odd dimensions, respectively; and where in the second
term we have used $|z|^2 = m^2\sigma^{\rho}\sigma_{\rho}$. To
reiterate, $k_d=(d-2)/2$ for $d$ even and $k_d=(d-3)/2$ for $d$
odd. While at this stage the even- and odd-dimensional equations appear
to have the same form (with the simple replacement $n\to
n-\tfrac{1}{2}$), it is clear from Eqs.\
(\ref{Eq:EvenBesselSmallArg})-(\ref{Eq:OddBesselSmallArg}) that the
end result is not the same.  In particular, as is known, the
even-dimensional result contains a logarithmic divergence while the
odd dimensional result does not.

The preceding equations are covariant expressions that isolate the
divergences in a generic $d$-dimensional spacetime.  To perform any
meaningful subtraction of these divergences from the Green's function,
these terms must be expressed in a form commensurate with Eqs.\
(\ref{Eq:GreenFuncZeroTemp})-(\ref{Eq:GreenFuncNonZeroTemp}).  In
particular, it would be nice if these terms could be expressed either
as an integral over $\omega$ or as a sum over $n$.  It will be shown
that useful integral and sum representations compatible with Eqs.\
(\ref{Eq:GreenFuncZeroTemp})-(\ref{Eq:GreenFuncNonZeroTemp}) can be
found for even-dimensional spacetimes.  At this time, however, a
correspondingly suitable expression for use with odd-dimensional
spacetimes remains elusive.  Consequently, in what follows we
primarily address renormalization with respect to even-dimensional
spacetimes.

Unfortunately it does not seem to be possible to obtain a simple,
compact, general expression as an integral over $\omega$ or sum over
$n$.  The first problem is that, while the second term of Eqs.\
(\ref{Eq:EvenCovariantRenormTerms})-(\ref{Eq:OddCovariantRenormTerms})
may simply be finite, as is the case for four dimensions, this is not
generally true for higher dimensions, as will be shown explicitly for
the six dimensional case below.  These additional divergent terms may
be addressed by Howard's method \cite{Howard:1984qp}, described in
Appendix \ref{App:HowardIdentities} and used below.

As for the first term, we may proceed a little further but must use
some care.  Recall that the physical parameter approaching zero is
$\varepsilon = \tau-\tau'$, then $z$ must be expanded in powers of
$\varepsilon$ with the end result that $z^2=-2m^2\sum_{n=1}^{\infty}
c_{2n}\varepsilon^{2n}$ for some $r$-dependent coefficients $c_{2n}$.
The $c_{2n}$ are combinations of the metric functions $f$ and $h$, and
their derivatives. Expanding $z^{-2n}$ in powers of $\varepsilon$ one
gets a series of terms proportional to $\varepsilon^{-n},
\varepsilon^{-n+1}, \dots \varepsilon^{-1}$ plus a constant term.
This means that
\begin{equation}
 (2m^2)^{\nu}|z|^{-\nu}K_{\nu}(|z|) =
 \frac{(2m^2)^{\nu}}{(m\varepsilon\sqrt{f})^{\nu}}
K_{\nu}(m\varepsilon\sqrt{f})
 + \mbox{Extra Terms}.
\end{equation}
The extra terms, which will be denoted $E_{\nu}$, must be determined
for each $\nu$, and so far a compact expression giving the extra terms
for a given $\nu$ is unavailable, but may possibly be found from a
lengthy exercise in combinatorial gymnastics.  The extra terms for
the first few integral and half-integral $\nu$ are presented in Tables
\ref{Tab:ExtrasGeneral} and expressed in terms of the coefficients
$c_{2n}$.
\begin{table}
\begin{centering}
\begin{tabular}{r|l} 
 $\nu$ & \mbox{Extra Terms}\\
 \hline
 0 & 0 \\
 $\frac{1}{2}$ & 0 \\
 1 & $-\frac{c_4}{c_2^2}$ \\
 $\frac{3}{2}$ & $-\frac{3}{2}\sqrt{\frac{\pi}{2c_2}}
\frac{c_4}{\varepsilon  c_2^2}$ \\
 2 & $-\frac{4 c_4}{\varepsilon ^2 c_2^3}+\frac{m^2 c_4}{c_2^2}
+\frac{2}{c_2^4}(3 c_4^2-2 c_2c_6)$ \\
 $\frac{5}{2}$ & $-\frac{3}{2}\sqrt{\frac{\pi}{2c_2}}
\left[\frac{5c_4}{\varepsilon ^3 c_2^3}-\frac{1}{\varepsilon}
\left(\frac{m^2c_4}{ c_2^2}+\frac{5}{4c_2^4}
\left(7c_4^2-4c_2c_6\right)\right)\right]$ \\
 3 & $-\frac{24 c_4}{\varepsilon ^4 c_2^4} + 
\frac{4}{\varepsilon ^2 c_2} 
\left(\frac{m^2 c_4}{c_2^2} + \frac{6}{c_2^4}(2c_4^2-c_2c_6\right)
-\frac{m^2}{2} 
\left(\frac{m^2c_4}{c_2^2}+\frac{4}{c_2^4}(3c_4^2 - 2c_2 c_6)
\right) -\frac{8}{c_2^6} 
\left(10 c_4^3-12 c_2 c_4 c_6+3 c_2^2 c_8\right)$\\
 $\frac{7}{2}$ & $-\frac{3}{2}\sqrt{\frac{\pi}{2c_2}}
\left[\frac{35c_4}{\varepsilon^5 c_2^4}-\frac{1}{\varepsilon^3}
\left(\frac{5 m^2 c_4}{ c_2^3}+\frac{35}{4c_2^5}
\left(9c_4^2-4c_2c_6\right)\right)+\frac{1}{\varepsilon}
\left(\frac{m^4c_4}{2c_2^2}+\frac{5m^2}{4c_2^4}(7c_4^2-4c_2c_6)+
\frac{35}{8 c_2^6} \left(33 c_4^3-36 c_2 c_4 c_6+8 c_2^2 c_8\right)
\right)\right]$\\
 4 & $-\frac{192 c_4}{\varepsilon ^6 c_2^5} + 
\frac{24}{\varepsilon ^4 c_2^2}\left(\frac{m^2c_4}{c_2^2} + 
\frac{4}{c_2^4}\left(5c_4^2-2c_2 c_6\right)\right) - 
\frac{2}{\varepsilon^2 c_2}\left[m^2\left(\frac{m^2 c_4^2}{c_2^2}+
\frac{12}{c_2^4}(2c_4^2-c_2c_6)\right)+\frac{96}{c_2^6} 
\left(5 c_4^3-5 c_2 c_4 c_6+c_2^2 c_8\right) \right]$ \\& $+ 
\frac{m^2}{3}\left[\frac{m^2}{2}\left(\frac{m^2c_4}{c_2^2}+
\frac{6}{c_2^4}(3c_4^2 - 2c_2 c_6)\right) +\frac{24}{c_2^6} 
\left(10 c_4^3-12 c_2 c_4 c_6+3 c_2^2 c_8\right) \right] +
\frac{48}{c_2^8} \left(35 c_4^4-60 c_2 c_4^2 c_6+20 c_2^2 c_4 c_8+ 
10 c_2^2c_6^2 - 4c_2^3 c_{10}\right)$ \\
\end{tabular}
\caption{This table shows the extra terms generated by the modified
Bessel function $K_{\nu}(|z|)$ when one makes the replacement $|z|^2
\to 2m^2\sum_{n=1}^{\infty} c_{2n}\varepsilon^{2n}$.  Integer values
of $\nu$ are applicable to even-dimensional spacetimes, whereas half-integer
 values of $\nu$ are applicable to odd-dimensional spacetimes.}
\label{Tab:ExtrasGeneral}
\end{centering}
\end{table}

In practice the extra terms are straightforward to calculate using a
computer algebra system.  One simply takes the difference of
$(2m^2)^{\nu}|z|^{-\nu}K_{\nu}(|z|)$, with $|z|^2 \to
2m^2\sum_{n=1}^{\infty} c_{2n}\varepsilon^{2n}$ and expanded around
$\varepsilon=0$, and $(2m^2)^{\nu}|z|^{-\nu}K_{\nu}(|z|)$ with the
replacement $|z|^2\to 2m^2c_2\varepsilon^2$.  One important point that
should be noted is that finding the coefficient $c_{2n}$ in the
expansion of $|z|$ requires one to first calculate $\sigma^{\mu}$ to
order $\varepsilon^{2n-1}$. In Table \ref{Tab:SpecificExtras} the
extra terms are presented explicitly in terms of the metric functions
$f$ and $h$ for the first few integral and half-integral values of
$\nu$.  In four dimensions $k_d = 1$, in which case $\nu$ ranges from
0 to 1 and these extra terms contribute no new divergences.  In six
dimensions $k_d=2$, $\nu$ ranges from 0 to 2, so one extra divergent
term arises when $\nu=2$.  In eight dimensions the extra divergences
come from both $\nu=2$ and $\nu=3$.  Obviously a similar situation
occurs for odd dimensions.

\begin{table}[ht]
\begin{tabular}{r|l}
 $\nu$ & \mbox{Extra Terms} $(E_{\nu})$\\
 \hline
 0 & 0 \\
 $\frac12$ & 0 \\
 1 & $\frac{f'^2}{24f^2h}$ \\
 $\frac32$ & $\frac{\sqrt{\pi}f'^2}{16\varepsilon f^{5/2} h}$ \\
 2 & $\frac{f'^2}{f^2h}\left[\frac{m^2}{3}
\left(\frac{1}{m^2\varepsilon^2 f}-
\frac{1}{8}\right) + \frac{1}{60h}\left(\frac{23f'^2}{24f^2}
+\frac{f'h'}{2fh}-
\frac{f''}{f}\right)\right]$ \\
 $\frac52$ & $\frac{\sqrt{\pi}f'^2}{8f^{5/2} h}
\left(\frac{5}{\varepsilon ^3 f}
-\frac{1}{\varepsilon }\left(\frac{m^2}{2}-\frac{f'}{4f h}
\left(\frac{17 f'}{16f}+
\frac{h'}{2h}-\frac{f''}{f'}\right)\right)\right)$ \\
 3 & $\frac{f'^2}{f^2h}\left[m^4\left(\frac{4}{m^4\varepsilon ^4f^2}-
\frac{1}{3m^2\varepsilon ^2f}+\frac{1}{48}\right)+\frac{m^2}{5}
\frac{f'}{f h}\left(\frac{1}{m^2\varepsilon ^2f}\left(\frac{7f'}{6f}
+\frac{h'}{2 h}-\frac{f''}{f'}\right)-\frac{1}{12}\left(\frac{23 f'}{24f}
+\frac{ h'}{2 h}-\frac{f''}{f'}\right)\right) \right. $
\\ & $\left. \qquad +\frac{f'}{f^2h^2}\left(-\frac{f'}{12}
\left(\frac{23f'^2}{360f^2}+
\frac{f' h'}{5f h}+\frac{h'^2}{16 h^2}\right)+
\frac{f''}{6}\left(\frac{f'}{5f}+
\frac{h'}{8 h}-\frac{f''}{8f'}\right)\right)
\right]$\\
\end{tabular}
\caption{This table shows the extra terms generated by the modified
Bessel function $K_{\nu}(|z|)$ for the specific metric of Eq.\
(\ref{Eq:Metric}).}\label{Tab:SpecificExtras}
\end{table}
All these divergent terms lurking about within Eqs.\
(\ref{Eq:EvenCovariantRenormTerms})-(\ref{Eq:OddCovariantRenormTerms})
must now be expressed as an integral over $\omega$ or a sum over $n$,
commensurate with Eqs.\
(\ref{Eq:GreenFuncZeroTemp})-(\ref{Eq:GreenFuncNonZeroTemp}).  To this
end we use an integral representation of the modified Bessel function,
some identities proved by Howard \cite{Howard:1984qp}, and the Plana
sum formula
\cite{Dahlquist:1997i,Dahlquist:1999,Anderson:1990jh,Ghika:1977vq},
\begin{equation} \label{Eq:PlanaSumFormula}
\int_{j}^{\infty}f(n)dn = \sum_{n=j}^{\infty}f(n) - \frac{1}{2}f(j) -
i\int_0^{\infty}\frac{dt}{e^{2\pi t}-1}\left[ f(j+it)-f(j-it) \right],
\end{equation}
to convert between integrals and sums.

\subsection{DeWitt-Schwinger Renormalization in 
Even dimensions} \label{sec:EvenDimensions}

\subsubsection{Renormalization formulas at zero and 
nonzero temperatures} \label{sec:Generaltreatment2}

For $\nu$ an integer it is shown in Appendix \ref{App:IntegralRep}
that an integral representation of $K_{\nu}(z)$ for small $z$ is
\begin{equation} \label{Eq:IntegralRepresentation}
K_{\nu}(z) =
\frac{(-1)^{\nu}\sqrt{\pi}}{\Gamma(\nu+\tfrac{1}{2})}
\left(\frac{z}{2}\right)^{\nu}
\int_0^{\infty}dt\cos(zt)(t^2+1)^{\nu-1/2}.
\end{equation}

For $T=0$ one has to
connect Eq.\ (\ref{Eq:IntegralRepresentation}) 
with the $e^{i \omega \varepsilon}$ dependence of Eq.
(\ref{Eq:GreenFuncZeroTemp}). Consider the change of variables
$t=\omega/\sqrt{m^2 f}$, and $z=m\varepsilon\sqrt{f}$; then
\begin{equation} \label{Eq:KIntRep}
\frac{(2m^2)^{\nu}}{(m\varepsilon\sqrt{f})^{\nu}}
K_{\nu}(m\varepsilon\sqrt{f})
= \frac{\sqrt{\pi}}{(-f)^{\nu}
\Gamma(\nu+\tfrac{1}{2})}\int_0^{\infty} \cos(\varepsilon\omega)
(\omega^2+m^2f)^{\nu-1/2} d\omega.
\end{equation}
This result generalizes the integral representation found by Anderson
\cite[Eq.\ (3.4a) and (3.4b)]{Anderson:1990jh}.

For nonzero temperature $T$, one has 
to connect Eq.\ (\ref{Eq:IntegralRepresentation}) 
with the $e^{i \kappa\varepsilon n}$ dependence of Eq.
(\ref{Eq:GreenFuncNonZeroTemp}). We instead make the change of
variables $t = x\kappa/\sqrt{m^2f}$ to first obtain
\begin{equation} \label{Eq:KIntRep2}
\frac{(2m^2)^{\nu}}{(m\varepsilon\sqrt{f})^{\nu}}K_{\nu}(m\varepsilon\sqrt{f}) 
=
\frac{\kappa\sqrt{\pi}}{(-f)^{\nu}\Gamma(\nu+\tfrac{1}{2})}
\int_0^{\infty}
\cos(\kappa \varepsilon x) (\kappa^2x^2+m^2f)^{\nu-1/2} dx.
\end{equation}
The Plana sum formula, Eq.\ (\ref{Eq:PlanaSumFormula}), enables the
integral in this equation to be converted into a sum plus some
residues and is valid if the function $f$
satisfies three conditions:
(i) $f(\tau+it)$ is holomorphic for $\tau\geq j$ for any $t$,
(ii) $\lim_{t\to\infty} \left|f(\tau+it)\right|e^{-2\pi |t|} = 0$ 
uniformly for every $\tau\geq j$, and
(iii)  $\lim_{\tau\to\infty} \int_{-\infty}^{\infty} dt 
\left|f(\tau+it)\right|e^{-2\pi |t|} = 0$.
A naive application of the Plana sum formula would be to use
$j=0$, corresponding to the lower limit of integration in Eq.\
(\ref{Eq:KIntRep2}).  However, for $j=0$ the integrand of Eq.\
(\ref{Eq:KIntRep2}) is not holomorphic at $\tau=0$.  Consequently, one
must break up the integral into two parts
\begin{equation} \label{Eq:IntSeparation}
\int_0^{\infty}dx \cos(\kappa\varepsilon x)(\kappa^2
x^2+m^2f)^{\nu-1/2} = \int_0^1 dx \cos(\kappa\varepsilon x)(\kappa^2
x^2+m^2f)^{\nu-1/2} + \int_1^{\infty} dx \cos(\kappa\varepsilon
x)(\kappa^2 x^2+m^2f)^{\nu-1/2}.
\end{equation}

For the first integral $\cos(\kappa\varepsilon x)\approx 1$ in the
coincidence limit and the solution may be expressed as a
hypergeometric function depending on $\nu$ \cite[Eq.\
(2.271)]{Gradshteyn},
\begin{equation}
 \int_0^1 dx \cos(\kappa\varepsilon x)(\kappa^2
x^2+m^2f)^{\nu-1/2} = (m^2f)^{\nu-1/2} \phantom{}_2F_1
\left(\frac{1}{2},\frac{1}{2}-\nu,\frac{3}{2},-\frac{\kappa^2}{m^2f}\right).
\end{equation}
In general, this
hypergeometric function is equivalent to a polynomial in half integer
powers of $(\kappa^2+m^2f)$ plus a logarithmic term.  Applying the Plana sum
formula to the second integral gives 
\begin{multline} \label{Eq:PlanadIntegral}
\int_1^{\infty} \cos(\kappa\varepsilon x)(\kappa^2
x^2+m^2f)^{\nu-1/2} dx =\sum_{n=1}^{\infty}\cos(\kappa\varepsilon n)
\left(\kappa^2n^2+m^2f\right)^{\nu-\frac{1}{2}} - 
\frac{1}{2}(\kappa^2+m^2f)^{\nu-\frac{1}{2}} \\ - i\int_0^{\infty}
\frac{dt}{e^{2\pi
t}-1}\left\{\left[(1+it)^2\kappa^2+m^2f\right]^{\nu-1/2} -
\left[(1-it)^2\kappa^2+m^2f\right]^{\nu-1/2}\right\}.
\end{multline}
Putting this together, we have
\begin{multline}
 \frac{(2m^2)^{\nu}}{(m\varepsilon\sqrt{f})^{\nu}}
K_{\nu}(m\varepsilon\sqrt{f}) \\ 
= 
\frac{\kappa\sqrt{\pi}}{(-f)^{\nu}\Gamma(\nu+\frac{1}{2})}
\left\{\sum_{n=1}^{\infty}
\cos(\kappa\varepsilon n)\left(\kappa^2n^2+m^2f\right)^{\nu-\frac{1}{2}} - 
\frac{1}{2}(\kappa^2+m^2f)^{\nu-\frac{1}{2}} + (m^2f)^{\nu-\frac{1}{2}}
\phantom{}_2F_1
\left(\frac{1}{2},\frac{1}{2}-\nu,\frac{3}{2},-\frac{\kappa^2}{m^2f}\right)
\right. \\ \left. - i\int_0^{\infty}
\frac{dt}{e^{2\pi
t}-1}\left\{\left[(1+it)^2\kappa^2+m^2f\right]^{\nu-1/2} -
\left[(1-it)^2\kappa^2+m^2f\right]^{\nu-1/2}
\right\}\right\}.
\end{multline}
This result generalizes the sum representation found by Anderson 
\cite[Eq.\ (3.7a) and (3.7b)]{Anderson:1990jh}.
Finally, the renormalization terms for the $d$-dimensional spacetime 
of Eq.\ (\ref{Eq:Metric}) are
\begin{multline} \label{Eq:GdivZero}
 G_{\mathrm{div}}(x,x') = \frac{2}{(4\pi)^{d/2}}
\sum_{k=0}^{k_d}\Bigg[\frac{\left[a_k\right]\sqrt{\pi}}{(-f)^{\nu}\Gamma(\nu+
\frac{1}{2})}\int_0^{\infty}
\cos(\omega\varepsilon)(\omega^2+m^2f)^{\nu-1/2}d\omega  \\ 
+ [a_k]E_{\nu} + 
\sum_{n=1}^{\nu}\sum_{p=1}^{2n}\sum_{j=0}^{p}\frac{2^{2n-1}(-m^2)^{\nu-n}
\Gamma(n)}{\Gamma(\nu-n+1)}\frac{a^j_k\Delta^{1/2}_{p-j}}{(\sigma^{\rho}
\sigma_{\rho})^n}\Bigg]
\end{multline}
for the case of a scalar field at zero temperature $T=0$, and
\begin{multline} \label{Eq:GdivNonZero}
  G_{\mathrm{div}}(x,x') = \frac{2}{(4\pi)^{d/2}}\sum_{k=0}^{k_d}
\Bigg\{\frac{\left[a_k\right]\kappa\sqrt{\pi}}{(-f)^{\nu}
\Gamma(\nu+\frac{1}{2})} 
\Bigg[\sum_{n=1}^{\infty}\cos(\kappa\varepsilon n)
\left(\kappa^2n^2+m^2f\right)^{\nu-\frac{1}{2}} - 
\frac{1}{2}(\kappa^2+m^2f)^{\nu-\frac{1}{2}} \\ - i\int_0^{\infty}
\frac{dt}{e^{2\pi
t}-1}\left\{\left[(1+it)^2\kappa^2+m^2f\right]^{\nu-1/2} -
\left[(1-it)^2\kappa^2+m^2f\right]^{\nu-1/2}
\right\} \\ + (m^2f)^{\nu-\frac{1}{2}}\phantom{}_2F_1
\left(\frac{1}{2},\frac{1}{2}-\nu,\frac{3}{2},-\frac{\kappa^2}{m^2f}\right)
\Bigg]+ [a_k]E_{\nu} + \sum_{n=1}^{\nu}\sum_{p=1}^{2n}\sum_{j=0}^{p}
\frac{2^{2n-1}(-m^2)^{\nu-n}\Gamma(n)}{\Gamma(\nu-n+1)}
\frac{a^j_k\Delta^{1/2}_{p-j}}{(\sigma^{\rho}\sigma_{\rho})^n}  \Bigg\}
\end{multline}
for a scalar field at nonzero temperature $T>0$.

\subsubsection{Examples: $d=4$ and $d=6$}\label{sec:Examples}
The formulas given above provide simple expressions to calculate
the renormalization terms for the generic even-dimensional spacetime
of Eq.\ (\ref{Eq:Metric}).  Below we mention the case $d=4$ and study
more carefully the case $d=6$.  For any $d$-dimensional spacetime with
line element given by Eq.\ (\ref{Eq:Metric}), we generalize
$\sigma^{\mu}$ to \cite{ChristensenThesis,Anderson:1994hg}
\begin{subequations}
\begin{equation}
 \sigma^{\tau} = -\varepsilon +
\frac{\varepsilon^3}{24}\frac{(f')^2}{fh} +\frac{\varepsilon
^5}{120}\left(\frac{f'^4}{8f^2h^2}+\frac{3}{16}
\frac{\left(f'\right)^3h'}{fh^3}-\frac{3}{8}\frac{f'^2f''}{fh^2}\right)
+ O(\varepsilon^7)
\end{equation}
\begin{equation}
  \sigma^r = \frac{\varepsilon^2 f'}{4h} -
\frac{\varepsilon^4}{24}\left(-\frac{f'^2h'}{8h^3}+
\frac{f'f''}{4h^2}\right) + O(\varepsilon^6)
\end{equation}
\begin{equation}
 \sigma^{\theta_i} = 0 \quad  i=1\dots d-2.
\end{equation}
\end{subequations}
In applying Eqs.\ (\ref{Eq:GdivZero}) and (\ref{Eq:GdivNonZero}), we
use the values for $E_{\nu}$ as listed in Table
\ref{Tab:SpecificExtras}.

\vskip0.3cm
$\it d=4$:
\vskip0.2cm

It is straightforward to show that in four dimensions Eqs.\
(\ref{Eq:GdivZero}) and (\ref{Eq:GdivNonZero}) are identical to those
obtained by Anderson \cite{Anderson:1990jh,Anderson:1994hg} and used
by several subsequent authors.  After letting $\varepsilon \to 0$ we
find, for $T=0$,
\begin{equation}
 G_{\mathrm{div}}(x,x') =
 -\frac{1}{4\pi^2}\int_0^{\infty}d\omega\,
\left[\frac{1}{f}(\omega^2+m^2f)^{1/2}+\frac{1}{2}
\left(\xi-\frac{1}{6}\right)R\,
 (\omega^2+m^2f)^{-1/2}\right] -
 \frac{f'}{192fh}\left(\frac{4}{r}+\frac{2f''}{f'}-
\frac{2f'}{f}-\frac{h'}{h}\right).
\end{equation}
One may verify that expanding Eq.\ (\ref{Eq:GdivNonZero}) correctly 
reproduces the results obtained by Anderson for $T>0$.

\vskip0.3cm
$\it d=6$:
\vskip0.2cm

Another, less trivial, example can be given for a scalar field in six
dimensions. We consider three classes of spacetime: spherical, flat,
and hyperbolic, corresponding to $\mathfrak{K} = 1$, $0$, or $-1$,
respectively.  Consider first the last terms of Eqs.\
(\ref{Eq:GdivZero}) and (\ref{Eq:GdivNonZero}).  Using the values of
$E_{\nu}$ given in Table \ref{Tab:SpecificExtras} and calculating the
last sums, we find
\begin{equation}
 \sum_{k=0}^{k_d} \left[[a_k]E_{\nu} +
 \sum_{n=1}^{\nu}\sum_{p=1}^{2n}\sum_{j=0}^{p}
\frac{2^{2n-1}(-m^2)^{\nu-n}\Gamma(n)}{\Gamma(\nu-n+1)}
\frac{a^j_k\Delta^{1/2}_{p-j}}{(\sigma^{\rho}\sigma_{\rho})^n}\right]
 = -\frac{1}{\varepsilon^2}
 \frac{f'}{6f^2h}\left(\frac{8}{r}+\frac{2f''}{f'}-\frac{3f'}{f}-
\frac{h'}{h}\right)
 + C^{\mathfrak{K}}_6
\end{equation}
where
\begin{multline}
 C^{\mathfrak{K}}_6 = m^2 \frac{f'}{24fh}\left(\frac{8}{r}+\frac{2f''}{f'}
-\frac{2f'}{f}-\frac{h'}{h}\right)+\frac{f'}{fh^2}\Bigg\{\frac{2}{r^3}
\left[\left(\frac{1}{5}-\xi \right)-\mathfrak{K} h\left(\frac{1}{6}-
\xi \right)\right]
\\ +\frac{1}{r^2}\left[\mathfrak{K} h\left(\frac{1}{6}-\xi \right)
\left(\frac{f'}{f}+\frac{h'}{2h}-\frac{f''}{f'}\right)+\frac{f'}{12f}-
\frac{5h'}{2h}\left(\frac{4}{25}-\xi \right)+\frac{f''}{f'}
\left(\frac{1}{10}-\xi \right)\right]
\\ +\frac{1}{r}\left[-\frac{5f'^2}{6f^2}\left(\frac{9}{25}-\xi \right)-
\frac{f'h'}{3fh}\left(\frac{3}{5}-\xi \right)+\frac{5h'^2}{6h^2}
\left(\frac{3}{50}-\xi \right)+\frac{f''}{f'}\left(\frac{f'}{f}
\left(\frac{73}{180}-\xi \right)+\frac{h'}{3h}\left(\frac{1}{5}+
\xi \right)\right) \right.
\\ \left. -\frac{h''}{3h}\left(\frac{1}{20}-\xi \right)-\frac{f'''}{15f'}
\right)+\frac{f'}{f}\left(\frac{f'^2}{8f^2}\left(\frac{27}{40}-\xi \right)-
\frac{7f'h'}{48fh}\left(\frac{127}{210}-\xi \right)+\frac{5h'^2}{48h^2}
\left(\frac{63}{100}-\xi \right)\right]
\\ -\frac{f''}{f}\left(\frac{7f'}{24f}\left(\frac{127}{210}-\xi \right)+
\frac{5h'}{24h}\left(\frac{61}{100}-\xi \right)-\frac{f''}{12f'}
\left(\frac{11}{20}-\xi \right)\right)-\frac{f'}{12f}\left(\frac{13}{20}-
\xi \right)\left(\frac{h''}{2h}-\frac{f'''}{f'}\right)
\\ +\frac{7h'^3}{240h^3}-\frac{19f''h'^2}{480f'h^2}-\frac{13h'h''}{480h^2}+
\frac{f''h''}{60f'h}+\frac{h'f'''}{40f'h}+\frac{h'''}{240h}-
\frac{f^{(4)}}{120f'}\Bigg\}.
\end{multline}

Using Eq.\ (\ref{Eq:IntRepEpsilon}) for the integral representation of
$\varepsilon^{-2}$, the $T=0$ divergent terms in the limit
$\varepsilon\to 0$ are
\begin{multline}
 G_{\mathrm{div}}(x,x') = \frac{2}{(4\pi)^3}
\Bigg\{\int_0^{\infty}d\omega\left[\frac{4}{3f^2}(\omega^2+m^2f)^{3/2}+
\frac2f(\xi-\tfrac{1}{6})R(\omega^2+m^2f)^{1/2}+
[a_2](\omega^2+m^2f)^{-1/2} \right. \\ \left. + 
\frac{f'\omega}{6f^2h}\left(\frac{8}{r}+\frac{2f''}{f'}-
\frac{3f'}{f}-\frac{h'}{h}\right)\right] + C_6^{\mathfrak{K}} \Bigg\}
\end{multline}
where, for $d=6$,
\begin{multline} \label{Eq:a2d6}
[a_2]= \tfrac{1}{240h^2}\Bigg\{\tfrac{16}{r^4}(\mathfrak{K} h-1)
\left[\left(-41+420 \xi -1080 \xi ^2\right)+
\mathfrak{K}h\left(29-360 \xi +1080 \xi ^2\right)\right]
\\ +\tfrac{16}{r^3}\left(\tfrac{f'}{f}-\tfrac{h'}{h}\right)
\left[\left(29-290 \xi +720 \xi ^2\right)-5 k h\left(5-54 \xi +
144 \xi ^2\right)\right] 
\\ +\tfrac{4}{r^2}\left[10\mathfrak{K}(1-6\xi)^2 \tfrac{f'h}{f}
\left(\tfrac{f'}{f}+\tfrac{h'}{h}-2 \tfrac{f''}{f'}\right)+
\left(23-140 \xi +120 \xi ^2\right)\tfrac{f'^2}{f^2}-4\left(4-85 
\xi +330 \xi ^2\right) \tfrac{f'}{f} \tfrac{h'}{h} \right. \\ \left. +
5\left(-15+56 \xi +96 \xi ^2\right) \tfrac{h'^2}{h^2}+4
\left(1-40 \xi +180 \xi ^2\right) \tfrac{f''}{f}-40(-1+5 \xi ) 
\tfrac{h''}{h}\right]
\\ +\tfrac{4}{r} \left[\tfrac{f'}{f}\left(-\left(23-140 \xi +120 
\xi ^2\right) \tfrac{f'^2}{f^2}+\left(-33+140 \xi +120 \xi ^2\right) 
\tfrac{h'^2}{h^2}\right) \right. \\ +2\tfrac{f''}{f}\left(\left(21-130 
\xi +120 \xi ^2\right)\tfrac{f'}{f}-\left(-13+40 \xi +120 \xi ^2\right) 
\tfrac{h'}{h}\right) \\ \left. -4(1-5 \xi )\left(\tfrac{h'}{h}\left(6 
\tfrac{f'^2}{f^2} -14 \tfrac{ h'^2}{h^2}\right)-\tfrac{h''}{h}\left(4 
\tfrac{f'}{f}-13\tfrac{h' }{h}\right)+4\tfrac{f^{(3)}}{f}-2 
\tfrac{h^{(3)}}{h}\right)\right]
\\ +\tfrac{f'^2}{f^2} \left(\left(21-110 \xi +30 \xi ^2\right) 
\tfrac{f'^2}{f^2}-2 \left(-13+70 \xi -30 \xi ^2\right) \tfrac{f'}{f} 
\tfrac{h'}{h}+5 \left(5-26 \xi +6 \xi ^2\right) \tfrac{h'^2}{h^2}\right)  \\ 
-2\tfrac{f'f''}{f^2}\left(2\left(13-70 \xi +30 \xi ^2\right) \tfrac{f'}{f} +
5 \left(5-27 \xi +12 \xi ^2\right) \tfrac{h'}{h} -10 \left(1-6 \xi +6 
\xi ^2\right) \tfrac{f''}{f'}\right) \\ +2(1-5\xi )\left(14 \tfrac{f'}{f} 
\tfrac{h'^3}{h^3}-19\tfrac{h'^2}{h^2} \tfrac{f''}{f}-\tfrac{f'}{f}
\tfrac{h''}{h}\left(5 \tfrac{f'}{f} +13 \tfrac{h' }{h}\right)+
8\tfrac{f''}{f} \tfrac{h''}{h}+2\tfrac{f^{(3)}}{f}\left(5\tfrac{f'}{f}+
6\tfrac{h'}{h} \right)+2 \tfrac{f'}{f} \tfrac{h^{(3)}}{h}-
4 \tfrac{f^{(4)}}{f}\right)\Bigg\}.
\end{multline}

On the other hand, for nonzero temperature $T>0$, Eq.\
(\ref{Eq:SumRepEpsilon}) may be employed to find a sum representation
of $\varepsilon^{-2}$ and one may show that
\begin{multline}
G_{\mathrm{div}}(x,x') = \frac{2}{(4\pi)^3}
\Bigg\{\kappa\sum_{n=1}^{\infty} \left[\frac{4}{3f^2}(\kappa^2n^2+
m^2f)^{3/2} + \frac{2}{f}(\xi-\tfrac16)R(\kappa^2n^2+m^2f)^{1/2} + 
[a_2](\kappa^2n^2+m^2f)^{-1/2} \right. \\ \left. + 
\frac{n\kappa f'}{6f^2h}\left(\frac{8}{r}+\frac{2f''}{f'}-
\frac{3f'}{f}-\frac{h'}{h} \right) \right] + \ln\left(\frac{\kappa + 
\sqrt{\kappa^2+m^2f}}{\sqrt{m^2f}} \right)\left[\frac12 m^4 + 
(\xi-\frac16)m^2R + [a_2] \right] \\ - 
\frac{\kappa}{6f^2}(2\kappa^2-m^2f)(\kappa^2+m^2f)^{1/2} - 
\frac{\kappa}{2}[a_2](\kappa^2+m^2f)^{-1/2} + \frac{\kappa^2}{12} + 
C^{\mathfrak{K}}_6
\\ -\frac{4i\kappa}{3f^2}\int_{0}^{\infty} \frac{dt}{e^{2\pi
t}-1}\left\{\left[(1+it)^2\kappa^2+m^2f\right]^{3/2}-\left[(1-it)^2
\kappa^2+m^2f\right]^{3/2}\right\} \\ 
- \frac{2i\kappa}{f}\left(\xi-\tfrac{1}{6}\right)R\int_{0}^{\infty}
\frac{dt}{e^{2\pi
t}-1}\left\{\left[(1+it)^2\kappa^2+m^2f\right]^{1/2}-
\left[(1-it)^2\kappa^2+m^2f\right]^{1/2}\right\}
\\ - i\kappa[a_2]\int_{0}^{\infty} \frac{dt}{e^{2\pi
t}-1}\left\{\left[(1+it)^2\kappa^2+m^2f\right]^{-1/2}-
\left[(1-it)^2\kappa^2+m^2f\right]^{-1/2}\right\}\Bigg\}.
\end{multline}

These results are quite general and unwieldy, but expressions such as
Eq.\ (\ref{Eq:a2d6}), for $[a_2]$, may become quite simple for
particular situations.  For a minimally coupled scalar field in a 
six-dimensional, 
asymptotically anti-de Sitter Reissner-Nordstr\"om black
hole spacetime with
\begin{equation}
 f = \mathfrak{K} - \frac{\Lambda}{3}r^2 - \frac{M}{r^3} + \frac{Q^2}{r^6},
\end{equation}
$[a_2]$ reduces to
\begin{equation}
 [a_2] = \frac{4(14Q^2-5Mr^3)^2}{75r^{16}};
\end{equation}
a remarkably simple result (note that in six dimensions a 
minimally coupled field has $\xi = 1/5$).

Put in perspective, we have presented a formula for the renormalization
terms of $\langle\phi^2(x)\rangle$, which in the particular important
case of a minimally coupled scalar field in a $(d=6)$-dimensional
asymptotically anti-de Sitter Reissner-Nordstr\"om black hole
spacetime, yields an extremely simple compact formula.  The dimension
$d=6$ is of importance because it is the simplest even higher
dimension that can be made compatible with the extra large dimension
or brane world scenarios.

\section{Estimate of $\langle\phi^2(x)\rangle$ for Massive Fields} 
\label{sec:Estimates}
Using the DeWitt-Schwinger expansion we have isolated the
divergent terms in the coincidence limit.  These were precisely the
terms of the expansion up to $k=k_d$.  The DeWitt-Schwinger expansion
does not give the correct results for $\langle\phi^2(x)\rangle$, even
after removing the divergences, because the
expansion depends only on the local structure of the
spacetime whereas the true field modes also depend in part on the
global structure of the spacetime, for example, the effective potential
around a black hole.  However, when the field is massive enough, the
higher order terms in the expansion may be used as an approximation to
the renormalized value of $\langle\phi^2(x)\rangle$ such that
\begin{equation}
 \langle\phi^2(x)\rangle\approx \lim_{x'\to x}G_{\mathrm{ren}}(x,x') = 
\lim_{z\to 0} \frac{2}{(4\pi)^{d/2}}\sum_{k>k_d}\Delta^{1/2}
a_k(2m^2)^{\nu}|z|^{-\nu}K_{\nu}(|z|).
\end{equation}
For $k>k_d$, $\nu<0$, in which case $|z|^{-\nu}K_{\nu}(|z|)= 2^{-(\nu+1)}
\Gamma(-\nu) + O(z^1)$ for small $z$ in both even and odd 
dimensions.  
It follows that 
\begin{equation}
 \langle\phi^2(x)\rangle\approx \frac{1}{(4\pi)^{d/2}}
\sum_{k>k_d}[a_k]m^{2\nu}\Gamma(-\nu)
= \begin{cases}
   (2^d\pi^{d/2}m^2)^{-1}[a_{\frac{d}{2}}] + \cdots & d\, 
\mathrm{even}\,,\\
   (2^d\pi^{(d-1)/2}m)^{-1}[a_{\frac{d-1}{2}}] + \cdots & d\,
\mathrm{odd}\,,
  \end{cases}
\end{equation}
in the coincidence limit.

\vskip0.3cm
$d=4$:
\vskip0.2cm

\begin{figure}[ht]
\centering
\subfigure[]{
\includegraphics[scale=1]{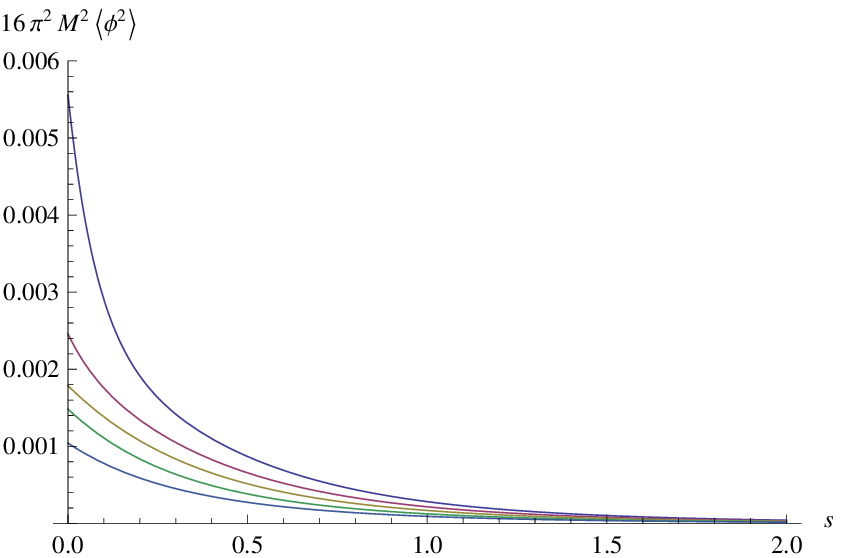}
\label{fig:4destimate}
}
\subfigure[]{
\includegraphics[scale=1]{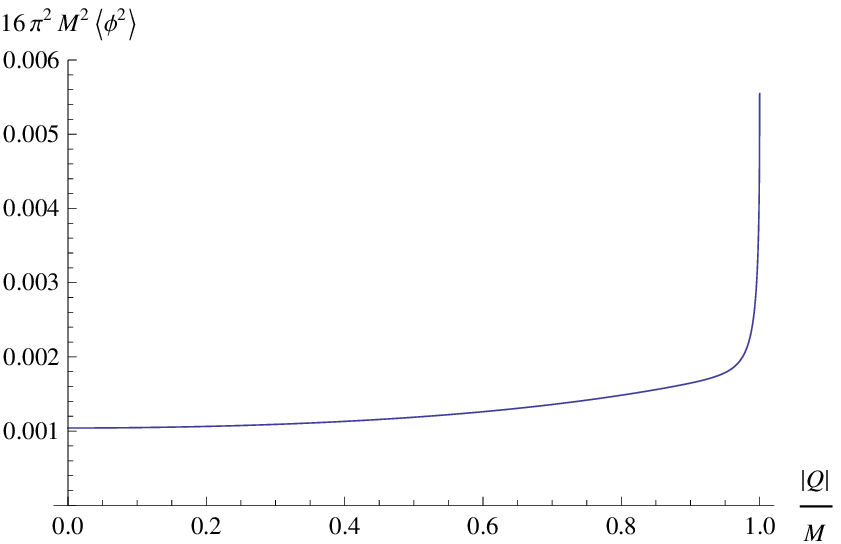}
\label{fig:4dPhiOnHorizon}
}
\label{fig:4destimates}
\caption{Plot of $\langle\phi^2(s)\rangle$ in a four-dimensional
 Reissner-Nordstr\"om black hole spacetime with
 $mM=2$. \subref{fig:4destimate} Near the horizon. From top to bottom
 the curves correspond to the cases $|Q|/M = 1.0,0.99,0.95,0.8,0.0$.
 The radial coordinate $s=r/M-1-\sqrt{1-(Q/M)^2}$ is zero at the
 horizon. \subref{fig:4dPhiOnHorizon} On the horizon as a function of
 $|Q|/M$.}
\end{figure}

For $d=4$ one finds $\langle\phi^2(x)\rangle_{d=4} \approx (4\pi
m)^{-2}[a_2]$.  This agrees exactly with the result reported by Anderson
\cite[Eq.\ (4.1)]{Anderson:1990jh}, but note that Anderson's equation
contains a typographical misprint in the first term,
which should be
$\tfrac16(\xi-\tfrac15)R_{;\rho}^{\phantom{;\rho};\rho}$.  This
misprint has no consequence since $R=0$ in the spacetime
considered by Anderson, but would be important elsewhere.  Calculating
the coefficient $[a_2]$ for a Reissner-Nordstr\"om black hole 
spacetime, for which
\begin{equation}
 f(r) = 1-\frac{2M}{r}+\frac{Q^2}{r^2},
\end{equation}
gives
\begin{equation}
 [a_2] = \frac{1}{45 r^8}(13 Q^4-24 M Q^2 r+12 M^2 r^2),
\end{equation}
leading to the near horizon behavior of
$\langle\phi^2(x)\rangle_{d=4}$ plotted in Fig.\ \ref{fig:4destimate}.
It may be seen that this correctly reproduces Fig.\ 3 of Ref.\
\cite{Anderson:1990jh}.  Figure \ref{fig:4dPhiOnHorizon} shows the
behavior of $\langle\phi^2(x)\rangle_{d=4}$ on the horizon as a
function of the charge-to-mass ratio, where it can be seen that the
value of $\langle\phi^2(x)\rangle_{d=4}$ increases rapidly to a finite
value as the black hole approaches extremality.

In \ Refs.\ \cite{Anderson:1990jh,Anderson:1994hg} it is emphasized
that the finite terms of the DeWitt-Schwinger expansion give a good
estimate for $\langle\phi^2(x)\rangle_{\mathrm{ren}}$ when $m M\gtrsim
1$, especially near the horizon. Since the horizon radius obeys
$r_h\sim M$ and the Compton wavelength associated to $m$ is
$\lambda\sim 1/m$, the rough inequality can be translated into
$r_h/\lambda\gtrsim1$.  This could have been expected on physical
grounds.  On one hand, particles with much longer wavelengths (lower
mass) are outside the validity of the approximation and it cannot give
good results.  On the other hand, since vacuum polarization happens
most intensely near the horizon, then particles with wavelengths on
the order of the horizon radius or less ($r_h/\lambda\gtrsim1$) are
well described by the approximation because they fit within the most
probable characteristic geometric length of the fully quantum
processes in a neighborhood of the horizon.

\vskip0.3cm
$\it d=5$:
\vskip0.2cm

\begin{figure}[ht]
\centering
\subfigure[]{
\includegraphics[scale=1]{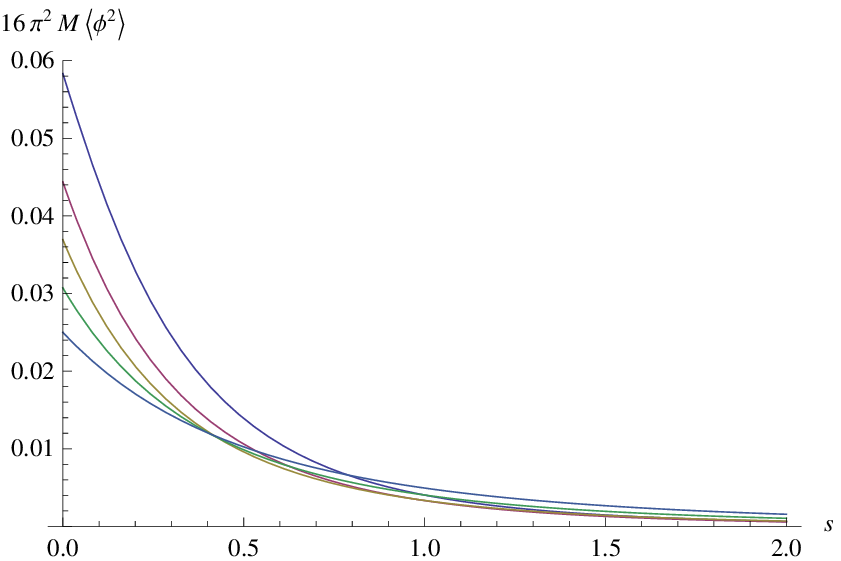}
\label{fig:5destimate0}
}
\subfigure[]{
\includegraphics[scale=1]{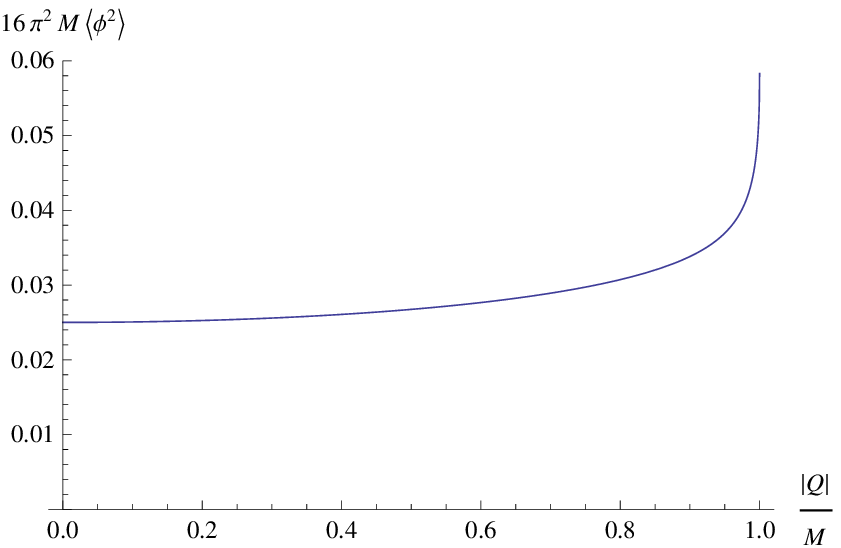}
\label{fig:5dPhiOnHorizon0}
}
\label{fig:5destimates0}
\caption{Plot of $\langle\phi^2(s)\rangle$ for a minimally coupled
field $(\xi=0)$ in a five-dimensional Reissner-Nordstr\"om black hole spacetime
with $mM=2$. \subref{fig:5destimate0} Near the horizon. From top
to bottom at $s=0$ the curves correspond to the cases $|Q|/M =
1.0,0.99,0.95,0.8,0.0$.  The radial coordinate
$s=r^2/M-1-\sqrt{1-(Q/M)^2}$ is zero at the horizon,
\subref{fig:5dPhiOnHorizon0} On the horizon as a function of $|Q|/M$.}
\end{figure}

\begin{figure}[ht]
\centering
\subfigure[]{
\includegraphics[scale=1]{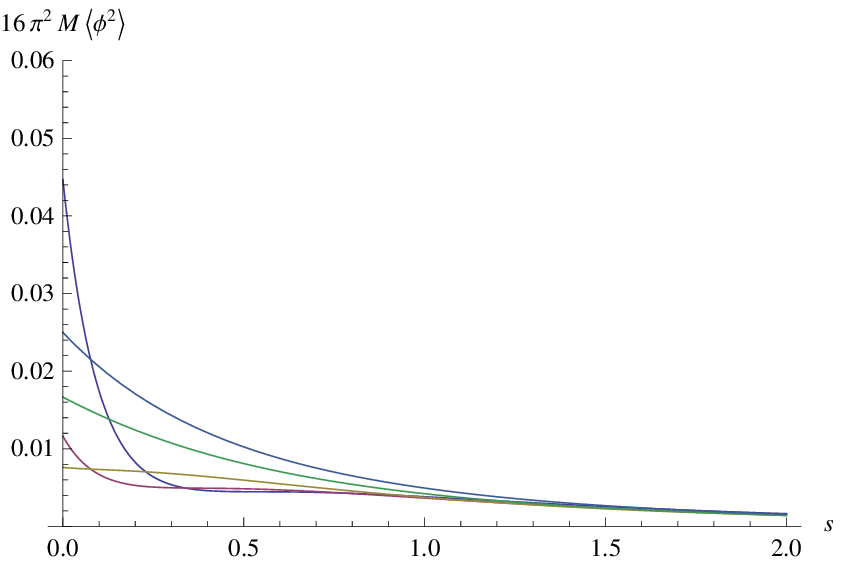}
\label{fig:5destimate1}
}
\subfigure[]{
\includegraphics[scale=1]{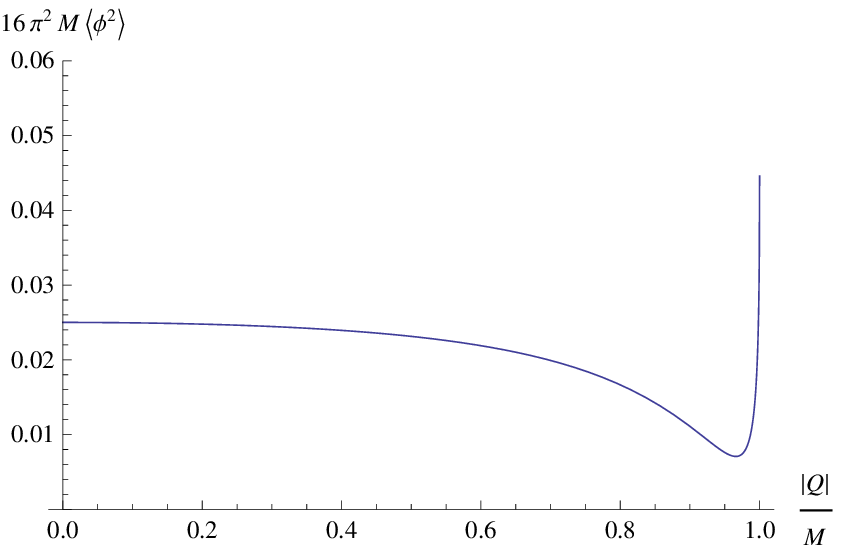}
\label{fig:5dPhiOnHorizon1}
}
\label{fig:5destimates1}
\caption[Optional caption for list of figures]{Plot of
$\langle\phi^2(s)\rangle$ for a conformally coupled field $(\xi=3/16)$
in a five-dimensional Reissner-Nordstr\"om black hole spacetime with
$mM=2$.  \subref{fig:5destimate0} Near the horizon. From top to bottom
at $s=0$ the curves correspond to the cases $|Q|/M =
1.0,0.0,0.8,0.99,0.95$.  The radial coordinate
$s=r^2/M-1-\sqrt{1-(Q/M)^2}$ is zero at the horizon,
\subref{fig:5dPhiOnHorizon0} On the horizon as a function of $|Q|/M$.}
\end{figure}

In five dimensions the metric function $f(r)$ for an asymptotically
flat Reissner-Nordstr\"om black hole is
\begin{equation} \label{Eq:5dimf}
 f(r) = 1-\frac{2M}{r^2}+\frac{Q^2}{r^4},
\end{equation}
from which we find
\begin{equation}
 [a_2] = \frac{1}{30 r^{12}}\left[48 M^2 r^4+Q^4 \left(-17+460 \xi +60 
\xi ^2\right)+24 Q^2 r^2 \left(M-30 M \xi +2 r^2 (-1+5 \xi )\right)\right].
\end{equation}
Notice that, unlike in four dimensions, in five dimensions $[a_2]$
depends on the coupling constant $\xi$.  From Eq.\ (\ref{Eq:5dimf}) it
is straightforward to locate the outer horizon at $r_h^2 =
M\left(1+\sqrt{1-(Q/M)^2}\right)$.  Letting $s=r^2-r_h^2$, the near horizon
behavior of $\langle\phi^2(x)\rangle_{d=5}$ is plotted in Figs.\
\ref{fig:5destimate0} and \ref{fig:5destimate1} for $\xi=0$ and
$\xi=3/16$ (conformal coupling) respectively.

The behavior of the conformally coupled field is quite interesting and
different from the behavior in four dimensions.  The value of
$\langle\phi^2(x)\rangle$ on the horizon is plotted in Figs.\
\ref{fig:5dPhiOnHorizon0} and \ref{fig:5dPhiOnHorizon1} for $\xi=0$
and $\xi=3/16$, respectively.  The main feature is that while
$\langle\phi^2(x)\rangle|_{x=x_h}$, the value on the horizon,
increases monotonically with $|Q|/M$ for the minimally coupled field
as for the four-dimensional case of Fig.\ \ref{fig:4dPhiOnHorizon},
for the conformally coupled field there is a minimum value which
occurs for a subextremal black hole with $|Q|/M\approx .97$.  These
results show that $\langle\phi^2(x)\rangle$ is well behaved near the
horizon as might be expected, but some interesting variation arises
for a non-minimally coupled field.

\section{Conclusions} \label{sec:Conclusions}

In this paper we have derived a compact expression for the
DeWitt-Schwinger renormalization terms in $d$ even dimensions.
Beginning with the general $d$-dimensional formula for the
DeWitt-Schwinger expansion, the divergent terms in the coincidence
limit were isolated by considering an expansion in $\sigma$.  From the
properties of modified Bessel functions a useful integral
representation for $K_{\nu}(z)$ in even dimensions was found in the
coincidence limit. This integral may be used for the case of a scalar
field at zero temperature without further modification.  For a scalar
field at nonzero temperature $T$ the Plana sum formula was used to
convert the integral into a sum plus residual terms.  The resulting
formulas, Eq.\ (\ref{Eq:GdivZero}) for a scalar field at zero
temperature, and Eq.\ (\ref{Eq:GdivNonZero}) for a scalar field at nonzero
temperature $T$, are given in a form compatible with calculations of
$\langle\phi^2(x)\rangle$ and $\langle T_{\mu\nu}(x)\rangle$ in static
spacetimes.  These formulas will be particularly useful for
calculating $\langle\phi^2(x)\rangle$ and $\langle T_{\mu\nu}(x)\rangle$ in
arbitrary black hole spacetimes of even dimension.  The formulas
found reproduce directly in the case $d=4$ the results obtained by
Anderson \cite{Anderson:1990jh}.  As a further example, the
renormalization terms were calculated for six dimensions. Christensen
remarked that calculating quantities such as $\langle
T_{\mu\nu}(x)\rangle$ in dimensions greater than four ``would be
extremely long and would probably have to be done on a computer''
\cite{Christensen:1978yd} mainly due to the complexity of the
renormalization problem.  While the derivation of Eqs.\
(\ref{Eq:GdivZero}) and Eq.\ (\ref{Eq:GdivNonZero}) did not require any
special computing power, it is certainly true that calculating
quantities such as $[a_2]$ and $C_d^{\mathfrak{K}}$ for a particular
spacetime would be quite lengthy and extremely prone to error without
the aid of a computer. Lastly, the finite terms of the
DeWitt-Schwinger expansion that are nonvanishing in the coincidence
limit may be used as an approximation to $\langle\phi^2(x)\rangle$.  It
is shown that this reduces to a sum over the DeWitt coefficients, and
is discussed in more detail in four and five dimensions.

As we have emphasized, in classical general relativity Einstein's
equations relate the spacetime curvature to the distribution of
classical matter as encoded in the stress-energy tensor.
Unfortunately the Universe is not so simple, being composed of
quantum, rather than classical, matter.  While some argue this
indicates the need for a complete quantum theory of gravity, a first
step, used here, is semiclassical general relativity, where the
stress tensor appearing in Einstein's equation is replaced by the
expectation value of the stress tensor of quantum fields.  Despite
this objection, semiclassical general relativity has provided us with
some impressive results and deep insight into the behavior of the
Universe, but calculating the renormalized expectation values of
$\langle\phi^2(x)\rangle$ and $\langle T_{\mu\nu}(x)\rangle$ is quite
difficult in curved spacetimes.  We have dealt with this difficulty
here for $d$-dimensional static spherical symmetric spacetimes in even
dimensions.  However, it is clear that, as it stands, the
semiclassical theory is inadequate as a complete theory of gravity at
the quantum level. By using the expectation value of the quantum
fields, information about fluctuations of the fields, a defining
characteristic of quantum field theory, is lost.  At the very minimum,
the semiclassical theory must be extended to incorporate some notion
of fluctuations, and work in this area is being done by several
authors; see for example
\cite{lrr-2008-3,ThompsonThesis,Thompson:2008pqa,Thompson:2008vi} and
references therein.

\begin{acknowledgments}
R.T.\ thanks Centro Multidisciplinar de Astrof\'{\i}sica -- CENTRA for
a grant through Funda\c c\~ao para a Ci\^encia e 
Tecnologia -- FCT (Portugal). This work was
partially funded by FCT, through Project No.\ PPCDT/FIS/57552/2004.
\end{acknowledgments}

\begin{appendix}
\section{Generalization of Howard's Identities} \label{App:HowardIdentities}
For $\varepsilon^{-2}$, Howard \cite{Howard:1984qp} proved that
\begin{equation} \label{Eq:SumRepEpsilon}
 \frac{1}{\varepsilon^2}=-\sum_{n=1}^{\infty} 
\kappa^2 n\cos(n\kappa\varepsilon) - \frac{\kappa^2}{12}
\end{equation}
for small, nonzero $\varepsilon$.  Using Howard's procedure for $p$ an
even integer this result is easily generalized, for the problem at
hand, to
\begin{equation}
 (m\varepsilon \sqrt{f})^{-p} = \left(\frac{i\kappa}{m\sqrt{f}}
\right)^p\frac{1}{p!}\left[p\sum_{n=1}^{\infty}n^{p-1}
\cos(n\kappa\varepsilon)+B_p\right],
\end{equation}
where $B_p$ are the Bernoulli numbers.
For $p$ an odd integer, the identity is generalized to
\begin{equation}
 (m\varepsilon\sqrt{f})^{-p}=-i\left(\frac{i\kappa}{m\sqrt{f}}
\right)^p\frac{p}{p!}\sum_{n=1}^{\infty}n^{p-1}\sin(n\kappa\varepsilon).
\end{equation}
An integral representation was proved by Anderson, et. al. 
\cite{Anderson:1994hg} by noting that
\begin{equation} \label{Eq:LnIntRep}
 \int_{\lambda}^{\infty} dt \frac{\cos(\varepsilon t)}{t} = -
\mathrm{ci}(\lambda\varepsilon) \sim -(\ln(\lambda\varepsilon) + 
\gamma) \quad \mathrm{as} \quad \varepsilon \to 0.
\end{equation}
By repeatedly taking the derivative of both sides of 
Eq.\ (\ref{Eq:LnIntRep}) and then letting $\lambda \to 0$, 
one finds
\begin{equation} \label{Eq:IntRepEpsilon}
 \frac{1}{\varepsilon^{2n}} = \frac{(-1)^n}{\Gamma(2n)} 
\int_0^{\infty}dt\, t^{2n-1}\cos(\varepsilon t)
\end{equation}
for even powers of $\varepsilon$, and
\begin{equation}
 \frac{1}{\varepsilon^{2n-1}} = 
\frac{(-1)^{n+1}}{\Gamma(2n+1)}\int_0^{\infty}dt \, 
t^{2n}\sin(\varepsilon t)
\end{equation}
for odd powers of $\varepsilon$.  In Appendix \ref{App:IntegralRep} 
we must evaluate
\begin{equation}
 \int_1^{\infty} dt\, t^{2n-1}\cos(\varepsilon t).
\end{equation}
This is done by writing
\begin{equation}
 \int_1^{\infty} dt\, t^{2n-1}\cos(\varepsilon t) =
 \int_0^{\infty}dt\, t^{2n-1}\cos(\varepsilon t) - \int_0^1 dt\,
 t^{2n-1}\cos(\varepsilon t) = \frac{(-1)^n
 \Gamma(2n)}{\varepsilon^{2n}} - \frac{1}{2n},
\end{equation}
where the $1/2n$ comes from expanding the solution of the second
integral near $\varepsilon =0$.

\section{Derivation of Eq.\ (\ref{Eq:IntegralRepresentation})} 
\label{App:IntegralRep}

To obtain formula (\ref{Eq:IntegralRepresentation}), 
begin with the recursion relation for the 
modified Bessel function,
\begin{equation} \label{Eq:recursion}
 K_{\nu+1}(z) = \frac{\nu}{z}K_{\nu}(z) - K_{\nu}'(z),
\end{equation}
and the integral representation for 
$K_0(z)$ \cite[Eq.\ (9.6.21)]{Abramowitz}
\begin{equation} \label{Eq:K0IntRep}
 K_0(z) = \int_0^{\infty}\frac{\cos(zt)dt}{(t^2+1)^{1/2}}.
\end{equation}
{}From the recursion relation $K_1(z) = -K_0'(z)$.  Taking the
derivative of Eq.\ (\ref{Eq:K0IntRep}) and integrating once by parts gives
\begin{equation} \label{Eq:K1integral}
K_1(z) = \int_0^{\infty}\frac{t\sin(zt)dt}{(t^2+1)^{1/2}} =
\lim_{s\to\infty}\sin(zt)(t^2+1)^{1/2}\Big|^{s}_0 -
z\int_0^{\infty}\cos(zt)(t^2+1)^{1/2}dt.
\end{equation}
Taking the limit $z\to 0$, the first term vanishes and only the
second term remains.  Applying the recursion relation again to find
$K_2(z)$, a fortuitous cancellation and another integration by parts
results in
\begin{equation}
K_2(z) = \frac{z^2}{3}\int_0^{\infty}\cos(zt)(t^2+1)^{3/2}dt
\approx
\frac{(-1)^{2}\sqrt{\pi}}{\Gamma(2+\tfrac{1}{2})}
\left(\frac{z}{2}\right)^2\int_0^{\infty}\cos(zt)(t^2+1)^{2-1/2}dt.
\end{equation}
By induction one is lead to
\begin{equation}
 K_{\nu}(z) = 
\frac{(-1)^{\nu}\sqrt{\pi}}{\Gamma(\nu+\tfrac{1}{2})}
\left(\frac{z}{2}\right)^2\int_0^{\infty}\cos(zt)(t^2+1)^{2-1/2}dt.
\end{equation}

One might worry that it is unreasonable to assume that the first term
on the right hand side of Eq.\ (\ref{Eq:K1integral}) vanishes, as the
limit $z\to 0$ should only be taken at the end of the calculation,
in which case the additional $z^{-1}$ multiplying $K_1(z)$ leaves
us with a problematic
$\lim_{s\to\infty}z^{-1}\sin(zt)(t^2-1)^{1/2}|_0^{s}$.

In fact, we can check that this integral representation reproduces the
correct limiting behavior of $z^{-\nu}K_{\nu}(z)$ when $z$ goes
linearly to 0.  First let
\begin{equation}
 \int_0^{\infty} dt \cos(zt)(t^2+1)^{\nu-1/2} = \int_0^{1}dt 
\cos(zt)(t^2+1)^{\nu-1/2}+\int_1^{\infty}dt \cos(zt)(t^2+1)^{\nu-1/2}.
\end{equation}
For $z$ in the coincidence limit we may expand the first integrand,
$\cos(zt)(t^2+1)^{\nu-1/2} \approx (t^2+1)^{\nu-1/2}$, and solve the
first integral
\begin{equation} \label{Eq:IntRepProof1}
 \int_0^{\infty} dt \cos(zt)(t^2+1)^{\nu-1/2} = 
\phantom{}_2F_1\left(\tfrac{1}{2},\tfrac{1}{2}-\nu,
\tfrac{3}{2},-1\right) + \int_1^{\infty}dt \cos(zt)(t^2+1)^{\nu-1/2}.
\end{equation}
For the second integral, expanding the integrand for $t>1$ gives
\begin{equation}
 \int_1^{\infty}dt \cos(zt)(t^2+1)^{\nu-1/2} = \Gamma\left(\nu+
\tfrac{1}{2}\right)\sum_{n=1}^{\infty}\frac{1}{\Gamma(n)\Gamma
\left(\nu-n+\tfrac{3}{2}\right)}\int_1^{\infty} dt \cos(zt)t^{2\nu-2n+1}.
\end{equation}
This expression may be analyzed with the use of some identities proved
by Anderson, et. al. \cite{Anderson:1994hg}, based on identities
proved by Howard \cite{Howard:1984qp}, and discussed further in
Appendix \ref{App:HowardIdentities}.  It follows from the results of
Appendix \ref{App:HowardIdentities} that
\begin{multline}
  \int_1^{\infty}dt \cos(zt)(t^2+1)^{\nu-1/2} =  
\Gamma\left(\nu+\tfrac{1}{2}\right)\left\{ \sum_{n=1}^{\nu} 
\frac{1}{\Gamma(n)\Gamma\left( \nu-n+\tfrac{3}{2}\right)}
\left[ \frac{(-1)^{\nu-n+1}\Gamma(2\nu-2n+2)}{z^{2\nu-2n+2}}-
\frac{1}{2(\nu-n+1)}\right] \right. \\ \left. - 
\frac{1}{\Gamma(\nu+1)\sqrt{\pi}}(\ln z + \gamma) + 
\sum_{\nu+2}^{\infty}\frac{1}{\Gamma(n)
\Gamma\left(\nu-n+\tfrac{3}{2}\right)}\int_1^{\infty}dt 
\cos(zt)t^{2\nu-2n+1}\right\}.
\end{multline}
Renaming indices on the last sum, rearranging terms slightly and 
combining with the first part of Eq.\ (\ref{Eq:IntRepProof1}),
\begin{multline}
  \int_0^{\infty}dt \cos(zt)(t^2+1)^{\nu-1/2} =  
\Gamma\left(\nu+\tfrac{1}{2}\right)\left\{ 
  \sum_{n=1}^{\nu} \frac{(-1)^{\nu-n+1}
\Gamma(2\nu-2n+2)}{\Gamma(n)
\Gamma\left(\nu-n+\tfrac{3}{2}\right)z^{2\nu-2n+2}}
  - \frac{1}{\Gamma(\nu+1)\sqrt{\pi}}\Bigg[ \ln z 
+ \gamma \right. 
  \\  \left. + \sum_{n=1}^{\nu}
\frac{\Gamma(\nu+1)\sqrt{\pi}}{2(\nu-n+1)
\Gamma(n)\Gamma\left(\nu-n+\tfrac{3}{2}\right)}
  - \sum_{n=1}^{\infty} \frac{\Gamma(\nu+1)
\sqrt{\pi}}{2n\Gamma(\nu+n+1)\Gamma\left(\tfrac{1}{2}-n\right)}
  - \frac{\Gamma(\nu+1)\sqrt{\pi}}{\Gamma\left(\nu+\tfrac12\right)} 
\phantom{}_2F_1\left(\tfrac{1}{2},\tfrac{1}{2}-
\nu,\tfrac{3}{2},-1\right) \Bigg] \right\}.
\end{multline}
The last sum may also be expressed in terms of a hypergeometric
function, and the limiting behavior is finally
\begin{multline} \label{Eq:IntRepProof}
 z^{-\nu}K_{\nu}(z) = \sum_{n=1}^{\nu} 
\frac{(-1)^{2\nu-n+1}\Gamma(2\nu-2n+2)\sqrt{\pi}}{2^{\nu}\Gamma(n)
\Gamma\left(\nu-n+\tfrac{3}{2}\right)z^{2\nu-2n+2}}
  - \frac{(-1)^{\nu}}{2^{\nu}\Gamma(\nu+1)}\Bigg[ \ln z + 
\gamma 
  + \sum_{n=1}^{\nu}\frac{\Gamma(\nu+1)\sqrt{\pi}}{2(\nu-n+1)
\Gamma(n)\Gamma\left(\nu-n+\tfrac{3}{2}\right)}
  \\ + \frac{\Gamma(\nu+1)}{4(\Gamma(\nu+2)}\phantom{}_3F_2
\left(1,1,\tfrac32,2,2+\nu,-1\right)
  - \frac{\Gamma(\nu+1)\sqrt{\pi}}{\Gamma\left(\nu+\tfrac12\right)} 
\phantom{}_2F_1\left(\tfrac{1}{2},\tfrac{1}{2}-
\nu,\tfrac{3}{2},-1\right) \Bigg] + O(z).
\end{multline}
This result does not look the same as the expansion given in Eq.\
(\ref{Eq:EvenBesselSmallArg}), but we checked the expansion explicitly
and believe the agreement is exact.  To have exact agreement, the last
three terms in brackets of Eq.\ (\ref{Eq:IntRepProof}) must be
identical to $-\sum_{n=1}^{\nu} n^{-1} - \ln 2$.  The two expressions
were evaluated numerically with a precision of 100 digits up to $\nu
=100$ and in each case were found to agree within the working
precision.  While this numerical evaluation does not constitute a
rigorous proof, it is a strong indication that the two expressions
agree exactly and therefore the integral representation is valid.

\end{appendix}


\begin{thebibliography}{71}
\expandafter\ifx\csname natexlab\endcsname\relax\def\natexlab#1{#1}\fi
\expandafter\ifx\csname bibnamefont\endcsname\relax
  \def\bibnamefont#1{#1}\fi
\expandafter\ifx\csname bibfnamefont\endcsname\relax
  \def\bibfnamefont#1{#1}\fi
\expandafter\ifx\csname citenamefont\endcsname\relax
  \def\citenamefont#1{#1}\fi
\expandafter\ifx\csname url\endcsname\relax
  \def\url#1{\texttt{#1}}\fi
\expandafter\ifx\csname urlprefix\endcsname\relax\def\urlprefix{URL }\fi
\providecommand{\bibinfo}[2]{#2}
\providecommand{\eprint}[2][]{\url{#2}}

\bibitem[{\citenamefont{Hawking}(1975)}]{Hawking:1974sw}
\bibinfo{author}{\bibfnamefont{S.~W.} \bibnamefont{Hawking}},
  \bibinfo{journal}{Commun. Math. Phys.} \textbf{\bibinfo{volume}{43}},
  \bibinfo{pages}{199} (\bibinfo{year}{1975}).

\bibitem[{\citenamefont{Boulware}(1975)}]{Boulware:1974dm}
\bibinfo{author}{\bibfnamefont{D.~G.} \bibnamefont{Boulware}},
  \bibinfo{journal}{Phys. Rev.} \textbf{\bibinfo{volume}{D11}},
  \bibinfo{pages}{1404} (\bibinfo{year}{1975}).

\bibitem[{\citenamefont{Unruh}(1976)}]{Unruh:1976db}
\bibinfo{author}{\bibfnamefont{W.~G.} \bibnamefont{Unruh}},
  \bibinfo{journal}{Phys. Rev.} \textbf{\bibinfo{volume}{D14}},
  \bibinfo{pages}{870} (\bibinfo{year}{1976}).

\bibitem[{\citenamefont{Hartle and Hawking}(1976)}]{Hartle:1976tp}
\bibinfo{author}{\bibfnamefont{J.~B.} \bibnamefont{Hartle}} \bibnamefont{and}
  \bibinfo{author}{\bibfnamefont{S.~W.} \bibnamefont{Hawking}},
  \bibinfo{journal}{Phys. Rev.} \textbf{\bibinfo{volume}{D13}},
  \bibinfo{pages}{2188} (\bibinfo{year}{1976}).

\bibitem[{\citenamefont{Birrell and Davies}(1982)}]{Birrell:1982ix}
\bibinfo{author}{\bibfnamefont{N.~D.} \bibnamefont{Birrell}} \bibnamefont{and}
  \bibinfo{author}{\bibfnamefont{P.~C.~W.} \bibnamefont{Davies}},
  \emph{\bibinfo{title}{{Quantum Fields in Curved Space}}}
  (\bibinfo{publisher}{Cambridge University Press},
  \bibinfo{address}{Cambridge}, \bibinfo{year}{1982}).

\bibitem[{\citenamefont{Fulling}(1989)}]{Fulling:1989nb}
\bibinfo{author}{\bibfnamefont{S.~A.} \bibnamefont{Fulling}},
  \emph{\bibinfo{title}{{Aspects of Quantum Field Theory in Curved
  Space-Time}}} (\bibinfo{publisher}{Cambridge University Press},
  \bibinfo{address}{Cambridge}, \bibinfo{year}{1989}).

\bibitem[{\citenamefont{Wald}(1994)}]{Wald:1995yp}
\bibinfo{author}{\bibfnamefont{R.~M.} \bibnamefont{Wald}},
  \emph{\bibinfo{title}{{Quantum Field Theory in Curved Space-Time and Black
  Hole Thermodynamics}}} (\bibinfo{publisher}{University of Chicago Press},
  \bibinfo{address}{Chicago}, \bibinfo{year}{1994}).

\bibitem[{\citenamefont{Schwinger}(1951)}]{Schwinger:1951nm}
\bibinfo{author}{\bibfnamefont{J.~S.} \bibnamefont{Schwinger}},
  \bibinfo{journal}{Phys. Rev.} \textbf{\bibinfo{volume}{82}},
  \bibinfo{pages}{664} (\bibinfo{year}{1951}).

\bibitem[{\citenamefont{DeWitt}(1965)}]{DeWitt:1965jb}
\bibinfo{author}{\bibfnamefont{B.~S.} \bibnamefont{DeWitt}},
  \emph{\bibinfo{title}{{Dynamical Theory of Groups and Fields}}}
  (\bibinfo{publisher}{Gordon \& Breach}, \bibinfo{address}{New York},
  \bibinfo{year}{1965}).

\bibitem[{\citenamefont{DeWitt}(1975)}]{DeWitt:1975ys}
\bibinfo{author}{\bibfnamefont{B.~S.} \bibnamefont{DeWitt}},
  \bibinfo{journal}{Phys. Rept.} \textbf{\bibinfo{volume}{19}},
  \bibinfo{pages}{295} (\bibinfo{year}{1975}).

\bibitem[{\citenamefont{Christensen}(1975)}]{ChristensenThesis}
\bibinfo{author}{\bibfnamefont{S.}~\bibnamefont{Christensen}}, Ph.D. thesis,
  \bibinfo{school}{University of Texas}, \bibinfo{address}{Austin, Texas}
  (\bibinfo{year}{1975}).

\bibitem[{\citenamefont{Christensen}(1976)}]{Christensen:1976vb}
\bibinfo{author}{\bibfnamefont{S.~M.} \bibnamefont{Christensen}},
  \bibinfo{journal}{Phys. Rev.} \textbf{\bibinfo{volume}{D14}},
  \bibinfo{pages}{2490} (\bibinfo{year}{1976}).

\bibitem[{\citenamefont{Christensen}(1978)}]{Christensen:1978yd}
\bibinfo{author}{\bibfnamefont{S.~M.} \bibnamefont{Christensen}},
  \bibinfo{journal}{Phys. Rev.} \textbf{\bibinfo{volume}{D17}},
  \bibinfo{pages}{946} (\bibinfo{year}{1978}).

\bibitem[{\citenamefont{Barvinsky and Vilkovisky}(1985)}]{Barvinsky:1985an}
\bibinfo{author}{\bibfnamefont{A.~O.} \bibnamefont{Barvinsky}}
  \bibnamefont{and} \bibinfo{author}{\bibfnamefont{G.~A.}
  \bibnamefont{Vilkovisky}}, \bibinfo{journal}{Phys. Rept.}
  \textbf{\bibinfo{volume}{119}}, \bibinfo{pages}{1} (\bibinfo{year}{1985}).

\bibitem[{\citenamefont{Brown and Collins}(1980)}]{Brown:1980qq}
\bibinfo{author}{\bibfnamefont{L.~S.} \bibnamefont{Brown}} \bibnamefont{and}
  \bibinfo{author}{\bibfnamefont{J.~C.} \bibnamefont{Collins}},
  \bibinfo{journal}{Ann. Phys.} \textbf{\bibinfo{volume}{130}},
  \bibinfo{pages}{215} (\bibinfo{year}{1980}).

\bibitem[{\citenamefont{Gilkey}(1975)}]{Gilkey:1975iq}
\bibinfo{author}{\bibfnamefont{P.~B.} \bibnamefont{Gilkey}},
  \bibinfo{journal}{J. Diff. Geom.} \textbf{\bibinfo{volume}{10}},
  \bibinfo{pages}{601} (\bibinfo{year}{1975}).

\bibitem[{\citenamefont{Avramidi}(1991)}]{Avramidi:1990je}
\bibinfo{author}{\bibfnamefont{I.~G.} \bibnamefont{Avramidi}},
  \bibinfo{journal}{Nucl. Phys.} \textbf{\bibinfo{volume}{B355}},
  \bibinfo{pages}{712} (\bibinfo{year}{1991}).

\bibitem[{\citenamefont{Amsterdamski et~al.}(1989)\citenamefont{Amsterdamski,
  Berkin, and O'Connor}}]{Amsterdamski:1989bt}
\bibinfo{author}{\bibfnamefont{P.}~\bibnamefont{Amsterdamski}},
  \bibinfo{author}{\bibfnamefont{A.~L.} \bibnamefont{Berkin}},
  \bibnamefont{and} \bibinfo{author}{\bibfnamefont{D.~J.}
  \bibnamefont{O'Connor}}, \bibinfo{journal}{Class. Quant. Grav.}
  \textbf{\bibinfo{volume}{6}}, \bibinfo{pages}{1981} (\bibinfo{year}{1989}).

\bibitem[{\citenamefont{Barvinsky et~al.}(1994)\citenamefont{Barvinsky, Gusev,
  Vilkovisky, and Zhytnikov}}]{Barvinsky:1994ic}
\bibinfo{author}{\bibfnamefont{A.~O.} \bibnamefont{Barvinsky}},
  \bibinfo{author}{\bibfnamefont{Y.~V.} \bibnamefont{Gusev}},
  \bibinfo{author}{\bibfnamefont{G.~A.} \bibnamefont{Vilkovisky}},
  \bibnamefont{and} \bibinfo{author}{\bibfnamefont{V.~V.}
  \bibnamefont{Zhytnikov}}, \bibinfo{journal}{J. Math. Phys.}
  \textbf{\bibinfo{volume}{35}}, \bibinfo{pages}{3543} (\bibinfo{year}{1994}),
  \eprint{arXiv:gr-qc/9404063}.

\bibitem[{\citenamefont{Kay and Wald}(1991)}]{Kay:1991}
\bibinfo{author}{\bibfnamefont{B.}~\bibnamefont{Kay}} \bibnamefont{and}
  \bibinfo{author}{\bibfnamefont{R.}~\bibnamefont{Wald}},
  \bibinfo{journal}{Phys. Rept.} \textbf{\bibinfo{volume}{207}},
  \bibinfo{pages}{49} (\bibinfo{year}{1991}).

\bibitem[{\citenamefont{Candelas}(1980)}]{Candelas:1980zt}
\bibinfo{author}{\bibfnamefont{P.}~\bibnamefont{Candelas}},
  \bibinfo{journal}{Phys. Rev.} \textbf{\bibinfo{volume}{D21}},
  \bibinfo{pages}{2185} (\bibinfo{year}{1980}).

\bibitem[{\citenamefont{Candelas and Howard}(1984)}]{Candelas:1984pg}
\bibinfo{author}{\bibfnamefont{P.}~\bibnamefont{Candelas}} \bibnamefont{and}
  \bibinfo{author}{\bibfnamefont{K.~W.} \bibnamefont{Howard}},
  \bibinfo{journal}{Phys. Rev.} \textbf{\bibinfo{volume}{D29}},
  \bibinfo{pages}{1618} (\bibinfo{year}{1984}).

\bibitem[{\citenamefont{Fawcett and Whiting}(1982)}]{Fawcett:1981fw}
\bibinfo{author}{\bibfnamefont{M.~S.} \bibnamefont{Fawcett}} \bibnamefont{and}
  \bibinfo{author}{\bibfnamefont{B.~F.} \bibnamefont{Whiting}}, in
  \emph{\bibinfo{booktitle}{Quantum structure of space and time}}, edited by
  \bibinfo{editor}{\bibfnamefont{M.~J.} \bibnamefont{Duff}} \bibnamefont{and}
  \bibinfo{editor}{\bibfnamefont{C.~J.} \bibnamefont{Isham}}
  (\bibinfo{publisher}{Cambridge University Press},
  \bibinfo{address}{Cambridge}, \bibinfo{year}{1982}), p. \bibinfo{pages}{131}.

\bibitem[{\citenamefont{Candelas and Jensen}(1986)}]{Candelas:1985ip}
\bibinfo{author}{\bibfnamefont{P.}~\bibnamefont{Candelas}} \bibnamefont{and}
  \bibinfo{author}{\bibfnamefont{B.~P.} \bibnamefont{Jensen}},
  \bibinfo{journal}{Phys. Rev.} \textbf{\bibinfo{volume}{D33}},
  \bibinfo{pages}{1596} (\bibinfo{year}{1986}).

\bibitem[{\citenamefont{Howard and Candelas}(1984)}]{Howard:1984qp}
\bibinfo{author}{\bibfnamefont{K.~W.} \bibnamefont{Howard}} \bibnamefont{and}
  \bibinfo{author}{\bibfnamefont{P.}~\bibnamefont{Candelas}},
  \bibinfo{journal}{Phys. Rev. Lett.} \textbf{\bibinfo{volume}{53}},
  \bibinfo{pages}{403} (\bibinfo{year}{1984}).

\bibitem[{\citenamefont{Howard}(1984)}]{Howard:1985yg}
\bibinfo{author}{\bibfnamefont{K.~W.} \bibnamefont{Howard}},
  \bibinfo{journal}{Phys. Rev.} \textbf{\bibinfo{volume}{D30}},
  \bibinfo{pages}{2532} (\bibinfo{year}{1984}).

\bibitem[{\citenamefont{Fawcett}(1983)}]{Fawcett:1983dk}
\bibinfo{author}{\bibfnamefont{M.~S.} \bibnamefont{Fawcett}},
  \bibinfo{journal}{Commun. Math. Phys.} \textbf{\bibinfo{volume}{89}},
  \bibinfo{pages}{103} (\bibinfo{year}{1983}).

\bibitem[{\citenamefont{Hawking}(1981)}]{Hawking:1980ng}
\bibinfo{author}{\bibfnamefont{S.~W.} \bibnamefont{Hawking}},
  \bibinfo{journal}{Commun. Math. Phys.} \textbf{\bibinfo{volume}{80}},
  \bibinfo{pages}{421} (\bibinfo{year}{1981}).

\bibitem[{\citenamefont{Frolov}(1982)}]{Frolov:1982pi}
\bibinfo{author}{\bibfnamefont{V.~P.} \bibnamefont{Frolov}},
  \bibinfo{journal}{Phys. Rev.} \textbf{\bibinfo{volume}{D26}},
  \bibinfo{pages}{954} (\bibinfo{year}{1982}).

\bibitem[{\citenamefont{Page}(1982)}]{Page:1982fm}
\bibinfo{author}{\bibfnamefont{D.~N.} \bibnamefont{Page}},
  \bibinfo{journal}{Phys. Rev.} \textbf{\bibinfo{volume}{D25}},
  \bibinfo{pages}{1499} (\bibinfo{year}{1982}).

\bibitem[{\citenamefont{Brown et~al.}(1986)\citenamefont{Brown, Ottewill, and
  Page}}]{Brown:1986jy}
\bibinfo{author}{\bibfnamefont{M.~R.} \bibnamefont{Brown}},
  \bibinfo{author}{\bibfnamefont{A.~C.} \bibnamefont{Ottewill}},
  \bibnamefont{and} \bibinfo{author}{\bibfnamefont{D.~N.} \bibnamefont{Page}},
  \bibinfo{journal}{Phys. Rev.} \textbf{\bibinfo{volume}{D33}},
  \bibinfo{pages}{2840} (\bibinfo{year}{1986}).

\bibitem[{\citenamefont{Zannias}(1984)}]{Zannias:1984tb}
\bibinfo{author}{\bibfnamefont{T.}~\bibnamefont{Zannias}},
  \bibinfo{journal}{Phys. Rev.} \textbf{\bibinfo{volume}{D30}},
  \bibinfo{pages}{1161} (\bibinfo{year}{1984}).

\bibitem[{\citenamefont{Frolov and Zelnikov}(1987)}]{Frolov:1987gw}
\bibinfo{author}{\bibfnamefont{V.~P.} \bibnamefont{Frolov}} \bibnamefont{and}
  \bibinfo{author}{\bibfnamefont{A.~I.} \bibnamefont{Zelnikov}},
  \bibinfo{journal}{Phys. Rev.} \textbf{\bibinfo{volume}{D35}},
  \bibinfo{pages}{3031} (\bibinfo{year}{1987}).

\bibitem[{\citenamefont{Frolov and Thorne}(1989)}]{Frolov:1989jh}
\bibinfo{author}{\bibfnamefont{V.~P.} \bibnamefont{Frolov}} \bibnamefont{and}
  \bibinfo{author}{\bibfnamefont{K.~S.} \bibnamefont{Thorne}},
  \bibinfo{journal}{Phys. Rev.} \textbf{\bibinfo{volume}{D39}},
  \bibinfo{pages}{2125} (\bibinfo{year}{1989}).

\bibitem[{\citenamefont{Elster}(1984)}]{Elster:1984hu}
\bibinfo{author}{\bibfnamefont{T.}~\bibnamefont{Elster}},
  \bibinfo{journal}{Class. Quant. Grav.} \textbf{\bibinfo{volume}{1}},
  \bibinfo{pages}{43} (\bibinfo{year}{1984}).

\bibitem[{\citenamefont{Frolov and Zelnikov}(1984)}]{Frolov:1984ra}
\bibinfo{author}{\bibfnamefont{V.~P.} \bibnamefont{Frolov}} \bibnamefont{and}
  \bibinfo{author}{\bibfnamefont{A.~I.} \bibnamefont{Zelnikov}},
  \bibinfo{journal}{Phys. Rev.} \textbf{\bibinfo{volume}{D29}},
  \bibinfo{pages}{1057} (\bibinfo{year}{1984}).

\bibitem[{\citenamefont{Jensen and Ottewill}(1989)}]{Jensen:1988rh}
\bibinfo{author}{\bibfnamefont{B.~P.}~\bibnamefont{Jensen}} \bibnamefont{and}
  \bibinfo{author}{\bibfnamefont{A.}~\bibnamefont{Ottewill}},
  \bibinfo{journal}{Phys. Rev.} \textbf{\bibinfo{volume}{D39}},
  \bibinfo{pages}{1130} (\bibinfo{year}{1989}).

\bibitem[{\citenamefont{Matyjasek}(1997)}]{Matyjasek:1996ih}
\bibinfo{author}{\bibfnamefont{J.}~\bibnamefont{Matyjasek}},
  \bibinfo{journal}{Phys. Rev.} \textbf{\bibinfo{volume}{D55}},
  \bibinfo{pages}{809} (\bibinfo{year}{1997}).

\bibitem[{\citenamefont{Anderson}(1989)}]{Anderson:1989vg}
\bibinfo{author}{\bibfnamefont{P.~R.} \bibnamefont{Anderson}},
  \bibinfo{journal}{Phys. Rev.} \textbf{\bibinfo{volume}{D39}},
  \bibinfo{pages}{3785} (\bibinfo{year}{1989}).

\bibitem[{\citenamefont{Anderson}(1990)}]{Anderson:1990jh}
\bibinfo{author}{\bibfnamefont{P.~R.} \bibnamefont{Anderson}},
  \bibinfo{journal}{Phys. Rev.} \textbf{\bibinfo{volume}{D41}},
  \bibinfo{pages}{1152} (\bibinfo{year}{1990}).

\bibitem[{\citenamefont{Anderson et~al.}(1993)\citenamefont{Anderson, Hiscock,
  and Samuel}}]{Anderson:1993if}
\bibinfo{author}{\bibfnamefont{P.~R.} \bibnamefont{Anderson}},
  \bibinfo{author}{\bibfnamefont{W.~A.} \bibnamefont{Hiscock}},
  \bibnamefont{and} \bibinfo{author}{\bibfnamefont{D.~A.}
  \bibnamefont{Samuel}}, \bibinfo{journal}{Phys. Rev. Lett.}
  \textbf{\bibinfo{volume}{70}}, \bibinfo{pages}{1739} (\bibinfo{year}{1993}).

\bibitem[{\citenamefont{Anderson et~al.}(1995)\citenamefont{Anderson, Hiscock,
  and Samuel}}]{Anderson:1994hg}
\bibinfo{author}{\bibfnamefont{P.~R.} \bibnamefont{Anderson}},
  \bibinfo{author}{\bibfnamefont{W.~A.} \bibnamefont{Hiscock}},
  \bibnamefont{and} \bibinfo{author}{\bibfnamefont{D.~A.}
  \bibnamefont{Samuel}}, \bibinfo{journal}{Phys. Rev.}
  \textbf{\bibinfo{volume}{D51}}, \bibinfo{pages}{4337} (\bibinfo{year}{1995}).

\bibitem[{\citenamefont{DeBenedictis}(1999)}]{DeBenedictis:1998be}
\bibinfo{author}{\bibfnamefont{A.}~\bibnamefont{DeBenedictis}},
  \bibinfo{journal}{Gen. Rel. Grav.} \textbf{\bibinfo{volume}{31}},
  \bibinfo{pages}{1549} (\bibinfo{year}{1999}), \eprint{arXiv:gr-qc/9804032}.

\bibitem[{\citenamefont{Piedra and de~Oca}(2008)}]{Piedra:2007yi}
\bibinfo{author}{\bibfnamefont{O.~P.~F.} \bibnamefont{Piedra}}
  \bibnamefont{and} \bibinfo{author}{\bibfnamefont{A.~C.~M.}
  \bibnamefont{de~Oca}}, \bibinfo{journal}{Phys. Rev.}
  \textbf{\bibinfo{volume}{D77}}, \bibinfo{pages}{024044}
  (\bibinfo{year}{2008}), \eprint{0707.0708}.

\bibitem[{\citenamefont{Sushkov}(2000)}]{Sushkov:2000me}
\bibinfo{author}{\bibfnamefont{S.~V.} \bibnamefont{Sushkov}},
  \bibinfo{journal}{Phys. Rev.} \textbf{\bibinfo{volume}{D62}},
  \bibinfo{pages}{064007} (\bibinfo{year}{2000}), \eprint{arXiv:gr-qc/0001058}.

\bibitem[{\citenamefont{Berej and Matyjasek}(2002)}]{Berej:2002xd}
\bibinfo{author}{\bibfnamefont{W.}~\bibnamefont{Berej}} \bibnamefont{and}
  \bibinfo{author}{\bibfnamefont{J.}~\bibnamefont{Matyjasek}},
  \bibinfo{journal}{Phys. Rev.} \textbf{\bibinfo{volume}{D66}},
  \bibinfo{pages}{024022} (\bibinfo{year}{2002}), \eprint{arXiv:gr-qc/0204031}.

\bibitem[{\citenamefont{Satz et~al.}(2005)\citenamefont{Satz, Mazzitelli, and
  Alvarez}}]{Satz:2004hf}
\bibinfo{author}{\bibfnamefont{A.}~\bibnamefont{Satz}},
  \bibinfo{author}{\bibfnamefont{F.~D.} \bibnamefont{Mazzitelli}},
  \bibnamefont{and} \bibinfo{author}{\bibfnamefont{E.}~\bibnamefont{Alvarez}},
  \bibinfo{journal}{Phys. Rev.} \textbf{\bibinfo{volume}{D71}},
  \bibinfo{pages}{064001} (\bibinfo{year}{2005}), \eprint{arXiv:gr-qc/0411046}.

\bibitem[{\citenamefont{Winstanley and Young}(2008)}]{Winstanley:2007tf}
\bibinfo{author}{\bibfnamefont{E.}~\bibnamefont{Winstanley}} \bibnamefont{and}
  \bibinfo{author}{\bibfnamefont{P.~M.} \bibnamefont{Young}},
  \bibinfo{journal}{Phys. Rev.} \textbf{\bibinfo{volume}{D77}},
  \bibinfo{pages}{024008} (\bibinfo{year}{2008}), \eprint{arXiv:0708.3820}.

\bibitem[{\citenamefont{Flachi and Tanaka}(2008)}]{Flachi:2008sr}
\bibinfo{author}{\bibfnamefont{A.}~\bibnamefont{Flachi}} \bibnamefont{and}
  \bibinfo{author}{\bibfnamefont{T.}~\bibnamefont{Tanaka}},
  \bibinfo{journal}{Phys. Rev.} \textbf{\bibinfo{volume}{D78}},
  \bibinfo{pages}{064011} (\bibinfo{year}{2008}), \eprint{0803.3125}.

\bibitem[{\citenamefont{Anderson et~al.}(2007)\citenamefont{Anderson, Mottola,
  and Vaulin}}]{Anderson:2007eu}
\bibinfo{author}{\bibfnamefont{P.~R.} \bibnamefont{Anderson}},
  \bibinfo{author}{\bibfnamefont{E.}~\bibnamefont{Mottola}}, \bibnamefont{and}
  \bibinfo{author}{\bibfnamefont{R.}~\bibnamefont{Vaulin}},
  \bibinfo{journal}{Phys. Rev.} \textbf{\bibinfo{volume}{D76}},
  \bibinfo{pages}{124028} (\bibinfo{year}{2007}), \eprint{arXiv:0707.3751}.

\bibitem[{\citenamefont{Popov and Zaslavskii}(2007)}]{Popov:2007ib}
\bibinfo{author}{\bibfnamefont{A.~A.} \bibnamefont{Popov}} \bibnamefont{and}
  \bibinfo{author}{\bibfnamefont{O.~B.} \bibnamefont{Zaslavskii}},
  \bibinfo{journal}{Phys. Rev.} \textbf{\bibinfo{volume}{D75}},
  \bibinfo{pages}{084018} (\bibinfo{year}{2007}), \eprint{arXiv:gr-qc/0703120}.

\bibitem[{\citenamefont{Frolov et~al.}(1989)\citenamefont{Frolov, Mazzitelli,
  and Paz}}]{Frolov:1989rv}
\bibinfo{author}{\bibfnamefont{V.~P.} \bibnamefont{Frolov}},
  \bibinfo{author}{\bibfnamefont{F.~D.} \bibnamefont{Mazzitelli}},
  \bibnamefont{and} \bibinfo{author}{\bibfnamefont{J.~P.} \bibnamefont{Paz}},
  \bibinfo{journal}{Phys. Rev.} \textbf{\bibinfo{volume}{D40}},
  \bibinfo{pages}{948} (\bibinfo{year}{1989}).

\bibitem[{\citenamefont{Casadio}(2004)}]{Casadio:2003jc}
\bibinfo{author}{\bibfnamefont{R.}~\bibnamefont{Casadio}},
  \bibinfo{journal}{Phys. Rev.} \textbf{\bibinfo{volume}{D69}},
  \bibinfo{pages}{084025} (\bibinfo{year}{2004}),
  \eprint{arXiv:hep-th/0302171}.

\bibitem[{\citenamefont{Decanini and Folacci}(2006)}]{Decanini:2005gt}
\bibinfo{author}{\bibfnamefont{Y.}~\bibnamefont{Decanini}} \bibnamefont{and}
  \bibinfo{author}{\bibfnamefont{A.}~\bibnamefont{Folacci}},
  \bibinfo{journal}{Phys. Rev.} \textbf{\bibinfo{volume}{D73}},
  \bibinfo{pages}{044027} (\bibinfo{year}{2006}), \eprint{gr-qc/0511115}.

\bibitem[{\citenamefont{Decanini and Folacci}(2008)}]{Decanini:2005eg}
\bibinfo{author}{\bibfnamefont{Y.}~\bibnamefont{Decanini}} \bibnamefont{and}
  \bibinfo{author}{\bibfnamefont{A.}~\bibnamefont{Folacci}},
  \bibinfo{journal}{Phys. Rev.} \textbf{\bibinfo{volume}{D78}},
  \bibinfo{pages}{044025} (\bibinfo{year}{2008}), \eprint{gr-qc/0512118}.

\bibitem[{\citenamefont{Decanini and Folacci}(2007)}]{Decanini:2007gj}
\bibinfo{author}{\bibfnamefont{Y.}~\bibnamefont{Decanini}} \bibnamefont{and}
  \bibinfo{author}{\bibfnamefont{A.}~\bibnamefont{Folacci}},
  \bibinfo{journal}{Class. Quant. Grav.} \textbf{\bibinfo{volume}{24}},
  \bibinfo{pages}{4777} (\bibinfo{year}{2007}), \eprint{0706.0691}.

\bibitem[{\citenamefont{Christensen and Fulling}(1977)}]{Christensen:1977jc}
\bibinfo{author}{\bibfnamefont{S.~M.} \bibnamefont{Christensen}}
  \bibnamefont{and} \bibinfo{author}{\bibfnamefont{S.~A.}
  \bibnamefont{Fulling}}, \bibinfo{journal}{Phys. Rev.}
  \textbf{\bibinfo{volume}{D15}}, \bibinfo{pages}{2088} (\bibinfo{year}{1977}).

\bibitem[{\citenamefont{Morgan et~al.}(2007)\citenamefont{Morgan, Thom,
  Winstanley, and Young}}]{Morgan:2007hp}
\bibinfo{author}{\bibfnamefont{D.}~\bibnamefont{Morgan}},
  \bibinfo{author}{\bibfnamefont{S.}~\bibnamefont{Thom}},
  \bibinfo{author}{\bibfnamefont{E.}~\bibnamefont{Winstanley}},
  \bibnamefont{and} \bibinfo{author}{\bibfnamefont{P.~M.} \bibnamefont{Young}},
  \bibinfo{journal}{Gen. Rel. Grav.} \textbf{\bibinfo{volume}{39}},
  \bibinfo{pages}{1719} (\bibinfo{year}{2007}), \eprint{arXiv:0705.1131}.

\bibitem[{\citenamefont{Herdeiro et~al.}(2008)\citenamefont{Herdeiro, Ribeiro,
  and Sampaio}}]{Herdeiro:2007eb}
\bibinfo{author}{\bibfnamefont{C.~A.~R.} \bibnamefont{Herdeiro}},
  \bibinfo{author}{\bibfnamefont{R.~H.} \bibnamefont{Ribeiro}},
  \bibnamefont{and} \bibinfo{author}{\bibfnamefont{M.}~\bibnamefont{Sampaio}},
  \bibinfo{journal}{Class. Quant. Grav.} \textbf{\bibinfo{volume}{25}},
  \bibinfo{pages}{165010} (\bibinfo{year}{2008}), \eprint{arXiv:0711.4564}.

\bibitem[{\citenamefont{Mamaev et~al.}(1976)\citenamefont{Mamaev, Mostepanenko,
  and Starobinsky}}]{Mamaev:1976je}
\bibinfo{author}{\bibfnamefont{S.~G.} \bibnamefont{Mamaev}},
  \bibinfo{author}{\bibfnamefont{V.~M.} \bibnamefont{Mostepanenko}},
  \bibnamefont{and} \bibinfo{author}{\bibfnamefont{A.~A.}
  \bibnamefont{Starobinsky}}, \bibinfo{journal}{JETP}
  \textbf{\bibinfo{volume}{43}}, \bibinfo{pages}{823} (\bibinfo{year}{1976}).

\bibitem[{\citenamefont{Bordag et~al.}(2001)\citenamefont{Bordag, Mohideen, and
  Mostepanenko}}]{Bordag:2001qi}
\bibinfo{author}{\bibfnamefont{M.}~\bibnamefont{Bordag}},
  \bibinfo{author}{\bibfnamefont{U.}~\bibnamefont{Mohideen}}, \bibnamefont{and}
  \bibinfo{author}{\bibfnamefont{V.~M.} \bibnamefont{Mostepanenko}},
  \bibinfo{journal}{Phys. Rept.} \textbf{\bibinfo{volume}{353}},
  \bibinfo{pages}{1} (\bibinfo{year}{2001}), \eprint{arXiv:quant-ph/0106045}.

\bibitem[{\citenamefont{Fetter and Walecka}(1971)}]{fetter}
\bibinfo{author}{\bibfnamefont{A.~L.} \bibnamefont{Fetter}} \bibnamefont{and}
  \bibinfo{author}{\bibfnamefont{J.~D.} \bibnamefont{Walecka}},
  \emph{\bibinfo{title}{{Quantum Theory of Many-Particle Systems}}}
  (\bibinfo{publisher}{McGraw-Hill}, \bibinfo{address}{New York},
  \bibinfo{year}{1971}).

\bibitem[{\citenamefont{Abramowitz and Stegun}(1964)}]{Abramowitz}
\bibinfo{author}{\bibfnamefont{M.}~\bibnamefont{Abramowitz}} \bibnamefont{and}
  \bibinfo{author}{\bibfnamefont{I.~A.} \bibnamefont{Stegun}},
  \emph{\bibinfo{title}{Handbook of Mathematical Functions with Formulas,
  Graphs, and Mathematical Tables}} (\bibinfo{publisher}{Dover},
  \bibinfo{address}{New York}, \bibinfo{year}{1964}).

\bibitem[{\citenamefont{Dahlquist}(1997)}]{Dahlquist:1997i}
\bibinfo{author}{\bibfnamefont{G.}~\bibnamefont{Dahlquist}},
  \bibinfo{journal}{BIT Numerical Mathematics} \textbf{\bibinfo{volume}{37}},
  \bibinfo{pages}{256} (\bibinfo{year}{1997}).

\bibitem[{\citenamefont{G.}(1999)}]{Dahlquist:1999}
\bibinfo{author}{\bibfnamefont{G.}~\bibnamefont{Dahlquist}}, 
\bibinfo{journal}{BIT  Numerical Mathematics} \textbf{\bibinfo{volume}{39}}, 
   \bibinfo{pages}{51}(\bibinfo{year}{1999}).

\bibitem[{\citenamefont{Ghika and Visinescu}(1978)}]{Ghika:1977vq}
\bibinfo{author}{\bibfnamefont{G.}~\bibnamefont{Ghika}} \bibnamefont{and}
  \bibinfo{author}{\bibfnamefont{M.}~\bibnamefont{Visinescu}},
  \bibinfo{journal}{Nuovo Cim.} \textbf{\bibinfo{volume}{A46}},
  \bibinfo{pages}{25} (\bibinfo{year}{1978}).

\bibitem[{\citenamefont{Gradshteyn and Ryzhik}(1994)}]{Gradshteyn}
\bibinfo{author}{\bibfnamefont{I.}~\bibnamefont{Gradshteyn}} \bibnamefont{and}
  \bibinfo{author}{\bibfnamefont{I.}~\bibnamefont{Ryzhik}},
  \emph{\bibinfo{title}{{Table of Integrals, Series, and Products}}}
  (\bibinfo{publisher}{Academic Press}, \bibinfo{address}{San Diego},
  \bibinfo{year}{1994}).

\bibitem[{\citenamefont{Thompson and
  Ford}(2008{\natexlab{a}})}]{Thompson:2008pqa}
\bibinfo{author}{\bibfnamefont{R.~T.} \bibnamefont{Thompson}} \bibnamefont{and}
  \bibinfo{author}{\bibfnamefont{L.~H.} \bibnamefont{Ford}},
  \bibinfo{journal}{Class. Quant. Grav.} \textbf{\bibinfo{volume}{25}},
  \bibinfo{pages}{154006} (\bibinfo{year}{2008}{\natexlab{a}}),
  \eprint{arXiv:0802.1546}.

\bibitem[{\citenamefont{Thompson and
  Ford}(2008{\natexlab{b}})}]{Thompson:2008vi}
\bibinfo{author}{\bibfnamefont{R.~T.} \bibnamefont{Thompson}} \bibnamefont{and}
  \bibinfo{author}{\bibfnamefont{L.~H.} \bibnamefont{Ford}},
  \bibinfo{journal}{Phys. Rev.} \textbf{\bibinfo{volume}{D78}},
  \bibinfo{pages}{024014} (\bibinfo{year}{2008}{\natexlab{b}}),
  \eprint{arXiv:0803.1980}.

\bibitem[{\citenamefont{Bei Lok~Hu}(2008)}]{lrr-2008-3}
\bibinfo{author}{\bibnamefont{B.~L.} \bibnamefont{Hu}} \bibnamefont{and} 
\bibinfo{author}{\bibnamefont{E.~} \bibfnamefont{Verdaguer}},
  \bibinfo{journal}{Living Reviews in Relativity} \textbf{\bibinfo{volume}{11}}
  (\bibinfo{year}{2008}),
  \urlprefix\url{http://www.livingreviews.org/lrr-2008-3}.

\bibitem[{\citenamefont{Thompson}(2008)}]{ThompsonThesis}
\bibinfo{author}{\bibfnamefont{R.~T.}~\bibnamefont{Thompson}}, Ph.D. thesis,
  \bibinfo{school}{Tufts University}, \bibinfo{address}{Medford, Massachusetts}
  (\bibinfo{year}{2008}).

\end{thebibliography}
\end{document}